\documentclass[prd,preprintnumbers,twocolumn,amsmath,nofootinbib,amssymb]{revtex4}
\usepackage{graphicx,color,dcolumn,booktabs,bm}
\usepackage{longtable,lscape}
\usepackage{txfonts}
\usepackage{overpic}
\usepackage{amssymb}
\usepackage{epstopdf}
\usepackage{indentfirst}
\usepackage{feynmf}   
\usepackage{slashed}  
\usepackage{cases}
\usepackage{color}
\usepackage{float}
\usepackage{multirow}
\usepackage{ulem}
\usepackage{graphicx,color,dcolumn,booktabs,bm}
\usepackage{epsfig,dsfont,amssymb,amsmath,amsfonts,amsbsy,mathrsfs}

\graphicspath{{Figures/}} %

\usepackage{hyperref}
\hypersetup{colorlinks,citecolor=blue,anchorcolor=red,menucolor=red, linkcolor=red,filecolor=red,runcolor=red,urlcolor=blue,frenchlinks=red}

\makeatletter
\@addtoreset{equation}{section}
\makeatother

\allowdisplaybreaks

\begin{document}

\title{Spectroscopic Properties of Double-Strangeness Molecular Tetraquarks}

\author{Fu-Lai Wang$^{1,2,3,5}$}
\email{wangfl2016@lzu.edu.cn}
\author{Si-Qiang Luo$^{1,2,3,4,5}$}
\email{luosq15@lzu.edu.cn}
\author{Ri-Qing Qian$^{1,2,3,5}$}
\email{qianrq@lzu.edu.cn}
\author{Xiang Liu$^{1,2,3,4,5}$\footnote{Corresponding author}}
\email{xiangliu@lzu.edu.cn}
\affiliation{$^1$School of Physical Science and Technology, Lanzhou University, Lanzhou 730000, China\\
$^2$Lanzhou Center for Theoretical Physics, Key Laboratory of Theoretical Physics of Gansu Province, Lanzhou University, Lanzhou 730000, China\\
$^3$Key Laboratory of Quantum Theory and Applications of MoE, Lanzhou University,
Lanzhou 730000, China\\
$^4$MoE Frontiers Science Center for Rare Isotopes, Lanzhou University, Lanzhou 730000, China\\
$^5$Research Center for Hadron and CSR Physics, Lanzhou University and Institute of Modern Physics of CAS, Lanzhou 730000, China}

\begin{abstract}
Inspired by recent advances in the study of the $K^{(*)} \bar K^{(*)}$ molecular tetraquarks and the $H$-dibaryon, we focus on the spectroscopic properties of the $\bar K^{(*)} \bar K^{(*)}$ systems, which exhibit exotic flavor quantum number of $ss\bar q \bar q$. A dynamical analysis is performed using the one-boson-exchange model to describe the effective interactions for these systems, accounting for both $S$-$D$ wave mixing and coupled-channel effects. By solving the coupled-channel Schr$\ddot{\rm o}$dinger equation, we identify the $I(J^P)=0(1^+)$ $\bar K \bar K^*$ and $I(J^P)=0(1^+)$ $\bar K^* \bar K^*$ states as the most likely candidates for double-strangeness molecular tetraquarks. Furthermore, we estimate their strong decay behaviors based on the effective Lagrangian approach, with several channels exhibiting considerable decay widths. Meanwhile, we investigate their magnetic moments and M1 radiative decay widths, shedding light on their inner structures, within the constituent quark model. Finally, we encourage experimentalists to focus on these predicted double-strangeness molecular tetraquark candidates, particularly in $B$ meson decays. Such efforts could pave the way for establishing the molecular tetraquark states in the light-quark sector.
\end{abstract}

\maketitle

\section{Introduction}\label{sec1}

The study of hadron spectroscopy offers a promising avenue for deepening our understanding of non-perturbative behavior of strong interaction, a rapidly advancing field at the precision frontier. Since the development of the quark model \cite{GellMann:1964nj,Zweig:1981pd}, extensive theoretical and experimental efforts have been devoted to identifying exotic hadronic candidates, which carry key insights into the nature of quantum chromodynamics (QCD) and present numerous challenges and opportunities for hadron physics. On the theory side, various exotic hadronic states have been proposed, including loosely molecular states, compact multiquark states, glueballs, hybrids, and other novel forms \cite{Liu:2013waa,Hosaka:2016pey,Chen:2016qju,Richard:2016eis,Lebed:2016hpi,Brambilla:2019esw,Liu:2019zoy,Chen:2022asf,Olsen:2017bmm,Guo:2017jvc,Meng:2022ozq}. In light of the growing number of newly observed hadronic states situated near two-hadron thresholds since the early 21st century \cite{Brambilla:2019esw}, the hadronic molecule has emerged as the most widely accepted framework for explaining these experimental findings \cite{Liu:2013waa,Hosaka:2016pey,Chen:2016qju,Richard:2016eis,Lebed:2016hpi,Brambilla:2019esw,Liu:2019zoy,Chen:2022asf,Olsen:2017bmm,Guo:2017jvc,Meng:2022ozq}, exemplified by discoveries such as $P_{c}(4312)$, $P_{c}(4440)$, $P_{c}(4457)$ \cite{Aaij:2019vzc}, $T_{cc}(3875)^+$ \cite{LHCb:2021vvq}, and others.

Since the Belle Collaboration's observation of the $X(3872)$ in 2003 \cite{Choi:2003ue}, numerous charmonium-like $XYZ$ states have been reported over the past 21 years \cite{Brambilla:2019esw}, including hidden charm molecular tetraquark candidates of the $D^{(*)} \bar D^{(*)}$-type \cite{Liu:2013waa,Hosaka:2016pey,Chen:2016qju,Richard:2016eis,Lebed:2016hpi,Brambilla:2019esw,Liu:2019zoy,Chen:2022asf,Olsen:2017bmm,Guo:2017jvc,Meng:2022ozq}. The $X(3872)$ remains the most well-known among them. However, a definitive confirmation of $D^{(*)} \bar D^{(*)}$ molecular tetraquarks is still pending \cite{Wang:2021aql}. In contrast, a significant breakthrough occurred in 2021 when the LHCb Collaboration discovered the double-charm tetraquark state $T_{cc}(3875)^+$ through an analysis of the $D^0D^0\pi^+$ invariant mass spectrum \cite{LHCb:2021vvq}. This state, with its exotic flavor quantum number of $cc \bar u \bar d$, clearly cannot be classified as a conventional hadron \cite{GellMann:1964nj,Zweig:1981pd}. Thus, the observation of the $T_{cc}(3875)^+$ provides strong evidence supporting the existence of the molecular tetraquark state in the charmed sector \cite{Manohar:1992nd,Ericson:1993wy,Tornqvist:1993ng,Janc:2004qn,Ding:2009vj,Molina:2010tx,Ding:2020dio,Li:2012ss,Xu:2017tsr,Liu:2019stu,Ohkoda:2012hv,Tang:2019nwv,Li:2021zbw,Chen:2021vhg,Ren:2021dsi,Xin:2021wcr,Chen:2021tnn,Albaladejo:2021vln,Dong:2021bvy,Baru:2021ldu,Du:2021zzh,Kamiya:2022thy,Padmanath:2022cvl,Agaev:2022ast,Ke:2021rxd,Zhao:2021cvg,Deng:2021gnb,Santowsky:2021bhy,Dai:2021vgf,Feijoo:2021ppq,Wang:2023ovj,Peng:2023lfw,Dai:2023cyo,Du:2023hlu,Kinugawa:2023fbf,Lyu:2023xro,Li:2023hpk,Dai:2023mxm,Wang:2022jop,Wu:2022gie,Ortega:2022efc,Jia:2022qwr,Praszalowicz:2022sqx,Chen:2022vpo,Lin:2022wmj,Cheng:2022qcm,Mikhasenko:2022rrl,He:2022rta}.

In light hadron spectroscopy, the search for $K^{(*)} \bar K^{(*)}$-type hidden strangeness molecular tetraquarks has drawn significant attention over the past several decades \cite{Guo:2017jvc,Close:2002zu}. During this period, considerable efforts have been made to interpret the $f_0(980)$ \cite{Astier:1967zz,Ammar:1968zur,Defoix:1968hip} and $f_1(1285)$ \cite{Gidal:1987bn,TPCTwoGamma:1988izb} as potential $K^{(*)} \bar K^{(*)}$ molecular tetraquark candidates \cite{Guo:2017jvc,Close:2002zu}. However, a definitive conclusion regarding their properties has not yet been reached. The lesson learned from the
establishment of the molecular tetraquark state in the charmed sector suggests that exotic flavor quantum number can provide a useful experimental handle for identifying and confirming exotic hadronic state \cite{Chen:2016qju}. Given this, both experimental and theoretical efforts should now focus on the $\bar K^{(*)} \bar K^{(*)}$-type double-strangeness molecular tetraquark candidates. The investigation of these double-strangeness tetraquark states is just as important as that of the double-charm tetraquark state $T_{cc}(3875)^+$ \cite{LHCb:2021vvq}, as it could yield critical insights into the formation of the molecular tetraquarks in the light-quark sector.

We now turn our attention to dibaryons with strangeness $S=-2$ \cite{Clement:2016vnl,Richard:2016eis}, another intriguing class of  exotic hadronic states that have long captured theoretical and experimental interest. Among these, Jaffe's 1976 prediction of a hexaquark state with the flavor content $uuddss$ and quantum numbers $I(J^P) = 0(0^+)$, known as the $H$-dibaryon, represents a bound state of the $\Lambda \Lambda$ system \cite{Jaffe:1976yi}. Despite significant attention from both experimental and theoretical perspectives, the existence of the $H$-dibaryon remains unresolved \cite{Takahashi:2001nm,E373KEK-PS:2013dfg,Belle:2013sba,Clement:2016vnl,Richard:2016eis}. Notably, far fewer hexaquark states have been observed compared to tetraquark states over the years \cite{Brambilla:2019esw}. In fact, the double-strangeness molecular tetraquark can be derived from the $H$-dibaryon by  substituting $qq \to \bar q$.

This study focuses on the spectroscopic properties of the $\bar K^{(*)} \bar K^{(*)}$ systems, characterized by an exotic flavor quantum number with the flavor content of $ss\bar q \bar q$. We conduct a dynamical analysis of these systems using the one-boson-exchange (OBE) model to describe their effective interactions. These interactions are then applied to search for loosely bound-state solutions by solving the coupled-channel Schr\"{o}dinger equation. This method allows us to predict the mass spectra of double-strangeness molecular tetraquarks, taking into account both $S$-$D$ wave mixing and coupled-channel effects.  Furthermore, we estimate the strong decay behaviors of the predicted double-strangeness molecular tetraquark candidates within the framework of the effective Lagrangian approach. Meanwhile, we explore the magnetic moments and M1 radiative decay widths of the predicted double-strangeness molecular tetraquark candidates based on the constituent quark model. The results of this study provide valuable information that could guide experimental efforts in the search for the $\bar K^{(*)} \bar K^{(*)}$ molecular tetraquark candidates. Successful identification of these states would offer important insights into the formation and establishment of the molecular tetraquarks in the light-quark sector.

The present work is organized as follows. In Sec. \ref{sec2}, we outline the detailed processes for deriving the effective interactions of the $\bar K^{(*)} \bar K^{(*)}$ systems. In Sec. \ref{sec3}, we discuss the bound-state properties of the $\bar K^{(*)} \bar K^{(*)}$ systems, and analyze the strong decay behaviors, magnetic moments, and M1 radiative decay widths of the $\bar K^{(*)} \bar K^{(*)}$ molecular tetraquark candidates. Finally, in Sec. \ref{sec4}, we provide a summary of the key finding of this work.

\section{Deriving the effective interactions of the $\bar K^{(*)} \bar K^{(*)}$ systems}\label{sec2}

A key topic of the present work is devoted to predict the mass spectra of  double-strangeness molecular tetraquarks, where the crucial step is the deduction of the effective interactions of the $\bar K^{(*)} \bar K^{(*)}$ systems. This section presents the detailed processes for deriving the effective interactions of the $\bar K^{(*)} \bar K^{(*)}$ systems. As is well known, the OBE model is a widely used tool for investigating the effective interactions between hadrons and identifying the observed new hadronic states under the hadronic molecule  framework \cite{Chen:2016qju}. Accordingly, the OBE model is employed in the present work to deduce the effective interactions of the $\bar K^{(*)} \bar K^{(*)}$ systems.

As listed in Ref. \cite{Chen:2016qju}, the OBE model enables the effective interactions between hadrons to be extracted through three principal steps. As the first step, we calculate the scattering amplitudes of the $\bar K^{(*)} \bar K^{(*)}\to \bar K^{(*)} \bar K^{(*)}$ processes, designated as $\mathcal{M}^{\bar K^{(*)} \bar K^{(*)} \to \bar K^{(*)} \bar K^{(*)}}(\bm{q})$. This is accomplished by utilising the effective Lagrangian approach, and the following relation is applicable:
\begin{eqnarray}
i\mathcal{M}^{\bar K^{(*)} \bar K^{(*)}\to \bar K^{(*)} \bar K^{(*)}}(\bm{q})=\sum_{E}i{\Gamma}^{\bar K^{(*)} \bar K^{(*)}E} D(q,m_{E}) i{\Gamma}^{\bar K^{(*)} \bar K^{(*)}E},
\end{eqnarray}
where ${\Gamma}^{\bar K^{(*)} \bar K^{(*)}E}$ and $D(q,m_{E})$ are the $\bar K^{(*)} \bar K^{(*)}E$ interaction vertices and the exchanged light boson propagator, respectively.

The $\bar K^{(*)} \bar K^{(*)}E$ interaction vertices can be extracted from the effective Lagrangians $\mathcal{L}_{\bar K^{(*)} \bar K^{(*)}E}$. As indicated in Refs. \cite{Liu:2020nil,Wang:2024ukc}, the following effective Lagrangians can be used to deduce the scattering amplitudes of the $\bar K^{(*)} \bar K^{(*)}\to \bar K^{(*)} \bar K^{(*)}$ processes, i.e.,
\begin{eqnarray}
\mathcal{L}_{\bar K^{(*)} \bar K^{(*)}\mathbb{P}}&=&-\frac{2ig^{\prime}}{f_{\pi}}v^{\alpha}\varepsilon_{\alpha\mu\nu\lambda}\widetilde K_b^{*\mu}\widetilde K_a^{*\lambda\dag}\partial^{\nu}{\mathbb{P}}_{ba}\nonumber\\
    &-&\frac{2g^{\prime}}{f_{\pi}}(\widetilde K_b^{*\mu}\widetilde K_a^{\dag}+\widetilde K_b \widetilde K_a^{*\mu\dag})\partial_{\mu}{\mathbb{P}}_{ba},\label{effectiveLagrangians1}\\
\mathcal{L}_{\bar K^{(*)} \bar K^{(*)} \sigma} &=&-2g_{s}^{\prime}{\widetilde K}_a \sigma {\widetilde K}_a^{\dag}+ 2g_{s}^{\prime} {\widetilde K}_{a\mu}^* \sigma {\widetilde K}_a^{*\mu\dag},\\
\mathcal{L}_{\bar K^{(*)} \bar K^{(*)}\mathbb{V}} &=&-\sqrt{2}\beta^{\prime} g_V \widetilde K_b \widetilde K_a^{\dag} v\cdot\mathbb{V}_{ba}+\sqrt{2}\beta^{\prime} g_V \widetilde K_{b\mu}^* \widetilde K_a^{*\mu\dag}v\cdot\mathbb{V}_{ba}\nonumber\\
    &-&2\sqrt{2}i\lambda^{\prime} g_V \widetilde K_b^{*\mu} \widetilde K_a^{*\nu\dag}\left(\partial_{\mu}\mathbb{V}_{\nu}-\partial_{\nu}\mathbb{V}_{\mu}\right)_{ba}\nonumber\\
    &-&2\sqrt{2}\lambda^{\prime} g_V v^{\lambda}\varepsilon_{\lambda\mu\alpha\beta}(\widetilde K_b \widetilde K_a^{*\mu\dag}+\widetilde K_b^{*\mu}\widetilde K_a^{\dag})\partial^{\alpha}\mathbb{V}^{\beta}_{ba}.\label{effectiveLagrangians2}
\end{eqnarray}
Here, $\widetilde K^{(*)}=(K^{(*)-}\,,\bar K^{(*)0})$, while the matrices employed to describe the light pseudoscalar mesons, denoted by the symbol ${\mathbb{P}}$, and the light vector mesons, represented by the symbol ${\mathbb{V}}$, are
\begin{eqnarray}
\left.\begin{array}{c}
{\mathbb{P}} = {\left(\begin{array}{ccc}
       \frac{\pi^0}{\sqrt{2}}+\frac{\eta}{\sqrt{6}} &\pi^+ &K^+\\
       \pi^-       &-\frac{\pi^0}{\sqrt{2}}+\frac{\eta}{\sqrt{6}} &K^0\\
       K^-         &\bar K^0   &-\sqrt{\frac{2}{3}} \eta     \end{array}\right)},\\
{\mathbb{V}} = {\left(\begin{array}{ccc}
       \frac{\rho^0}{\sqrt{2}}+\frac{\omega}{\sqrt{2}} &\rho^+ &K^{*+}\\
       \rho^-       &-\frac{\rho^0}{\sqrt{2}}+\frac{\omega}{\sqrt{2}} &K^{*0}\\
       K^{*-}         &\bar K^{*0}   & \phi     \end{array}\right)},
\end{array}\right.
\end{eqnarray}
respectively. Furthermore, the normalization relations for the strange mesons $\bar K$ and $\bar K^*$ in the effective Lagrangians ~(\ref{effectiveLagrangians1})-(\ref{effectiveLagrangians2}) are $\sqrt{m_{\bar K}}$ and
$\sqrt{m_{\bar {K}^{*}}}\epsilon^\mu$, respectively. In the static limit, the polarization vector of the strange meson $\bar K^*$, denoted by the variable $\epsilon ^\mu$, is given by $\epsilon_{0}^{\mu}= \left(0,0,0,-1\right)$  and $\epsilon_{\pm1}^{\mu}= (0,\,\pm{1}/{\sqrt{2}},\,{i}/{\sqrt{2}},\,0)$.

As a next step, we derive the effective interactions in the momentum space of the $\bar K^{(*)} \bar K^{(*)}\to \bar K^{(*)} \bar K^{(*)}$ processes, represented as ${V}^{\bar K^{(*)} \bar K^{(*)} \to \bar K^{(*)} \bar K^{(*)}}(\bm{q})$, which is achieved by employing the Breit approximation \cite{Berestetskii:1982qgu} and can be calculated using this equation:
\begin{eqnarray}
{V}^{\bar K^{(*)} \bar K^{(*)}\to \bar K^{(*)} \bar K^{(*)}}(\bm{q})=-\frac{\mathcal{M}^{\bar K^{(*)} \bar K^{(*)}\to \bar K^{(*)} \bar K^{(*)}}(\bm{q})} {4\sqrt{m_{\bar K^{(*)}}m_{\bar K^{(*)}}m_{\bar K^{(*)}}m_{\bar K^{(*)}}}}.
\end{eqnarray}

As a final step, we deduce the effective interactions in the coordinate space of the $\bar K^{(*)} \bar K^{(*)}\to \bar K^{(*)} \bar K^{(*)}$ processes, denoted as ${V}^{\bar K^{(*)} \bar K^{(*)}\to \bar K^{(*)} \bar K^{(*)}}(\bm{r})$. This is obtained by conducting the Fourier transformation for ${V}^{\bar K^{(*)} \bar K^{(*)}\to \bar K^{(*)} \bar K^{(*)}}(\bm{q})$ and ${F}_{\rm M}^2(q,\,m_{E})$, which is expressed as follows:
\begin{eqnarray}
&&{V}^{\bar K^{(*)} \bar K^{(*)}\to \bar K^{(*)} \bar K^{(*)}}(\bm{r})\nonumber\\ &&=\frac{1}{(2\pi)^3}\int{d^3\bm{q}}e^{i\bm{q}\cdot\bm{r}}{V}^{\bar K^{(*)} \bar K^{(*)}\to \bar K^{(*)} \bar K^{(*)}}(\bm{q}){F}_{\rm M}^2(q,\,m_{E}).
\end{eqnarray}
In consideration of the inner structures of the mesons, the monopole-type form factor ${F}_{\rm M}(q,\,m_{E}) = (\Lambda^2-m_{E}^2)/(\Lambda^2-q^2)$ was integrated at both $\bar K^{(*)} \bar K^{(*)}E$ interaction vertices \cite{Tornqvist:1993ng,Tornqvist:1993vu}. Here, $\Lambda$ is the cutoff parameter, $q$ and $m_{{E}}$ are the four momentum and the mass of the exchanged boson, respectively. This can be viewed as a means of offsetting the influence of the off-shell effects of the exchanged bosons, which was applied to successfully depict the bound-state properties of the deuteron \cite{Urey:1932gik}, $P_{c}(4312)$, $P_{c}(4440)$, $P_{c}(4457)$ \cite{Aaij:2019vzc}, and $T_{cc}(3875)^+$ \cite{LHCb:2021vvq} under the hadronic molecule  framework. In light of the masses of the deuteron \cite{Urey:1932gik}, $P_{c}(4312)$, $P_{c}(4440)$, $P_{c}(4457)$ \cite{Aaij:2019vzc}, and $T_{cc}(3875)^+$ \cite{LHCb:2021vvq} can be reproduced under the hadronic molecule picture when the cutoff value is approximately 1 GeV, it can be recommended that a bound state with a small binding energy and a large root-mean-square radius is the most likely candidate for the hadronic molecular state with a cutoff value around 1 GeV \cite{Chen:2016qju}.

As part of the derivation of the effective interactions of the $\bar K^{(*)} \bar K^{(*)}$ systems, it is essential to construct their wave functions. The flavor wave functions, represented by the notation $|I,I_{3}\rangle$, of the $\bar K \bar K$ and $\bar K^{*} \bar K^{*}$ systems are constructed as follows:
\begin{eqnarray}
|1, 1\rangle&=&\bar K^{(*)0} \bar K^{(*)0},\label{flavorwavefunctions1}\\
|1, 0\rangle&=&-\dfrac{1}{\sqrt{2}}\left(K^{(*)-} \bar K^{(*)0}+\bar K^{(*)0}K^{(*)-}\right),\\
|1, -1\rangle&=&K^{(*)-}K^{(*)-},\\
|0, 0\rangle&=&\dfrac{1}{\sqrt{2}}\left(K^{(*)-} \bar K^{(*)0}-\bar K^{(*)0}K^{(*)-}\right),\label{flavorwavefunctions2}
\end{eqnarray}
where $I$ and $I_3$ are the isospins and the isospin third components of the $\bar K^{(*)} \bar K^{(*)}$ systems, respectively. In accordance with the Einstein-Bose statistics, the permitted quantum numbers $I(J^P)$ of the $S$-wave $\bar K \bar K$ and $\bar K^{*} \bar K^{*}$ states are\footnote{In the present work, a two-body system comprising hadrons $A$ and $B$ is designated by the notation $[AB]$.}
\begin{eqnarray*}
[\bar K \bar K]:1(0^+)~~~{\rm and}~~~
[\bar K^{*} \bar K^{*}]:1(0^+)\,,0(1^+)\,, 1(2^+),
\end{eqnarray*}
respectively. In the present study, we discuss the impact of $S$-$D$ wave mixing and coupled-channel effects for the bound-state properties of the $\bar K^{(*)} \bar K^{(*)}$ systems. In light of the aforementioned considerations, the $S$- and $D$-wave channels $|^{2S+1}L_J\rangle$ for the $\bar K^{(*)} \bar K^{(*)}$ systems are
\begin{eqnarray*}
&&[\bar K \bar K][1(0^+)]:|{}^1\mathbb{S}_0\rangle,\\
&&[\bar K \bar K^*][0(1^+)]: |{}^3\mathbb{S}_1\rangle/|{}^{3}\mathbb{D}_1\rangle,~~
[\bar K \bar K^*][1(1^+)]: |{}^3\mathbb{S}_1\rangle/|{}^{3}\mathbb{D}_1\rangle,\\
&&[\bar K^* \bar K^*][1(0^+)]: |{}^1\mathbb{S}_0\rangle/|{}^{5}\mathbb{D}_0\rangle,~~
[\bar K^* \bar K^*][0(1^+)]: |{}^3\mathbb{S}_1\rangle/|{}^{3}\mathbb{D}_1\rangle,\\
&&[\bar K^* \bar K^*][1(2^+)]: |{}^5\mathbb{S}_2\rangle/|{}^{1}\mathbb{D}_2\rangle/|{}^{5}\mathbb{D}_2\rangle.
\end{eqnarray*}
To present the spin $S$, the orbital angular momentum $L$, and the total angular momentum $J$ of the $S$- and $D$-wave channels, we take the notation $|^{2S+1}L_J\rangle$. Besides, the spin-orbital wave functions $|{}^{2S+1}L_{J}\rangle$ of the $\bar K \bar K^{*}$ and $\bar K^{*} \bar K^{*}$ systems can be expressed as follows:
\begin{eqnarray}
&&|^{2S+1}L_J\rangle = \sum_{m,m_L}C^{J,M}_{1m,Lm_L}\epsilon_{m}^\mu Y_{L, m_L},\label{spin-orbitalwavefunctions1}\\
&&|^{2S+1}L_J\rangle = \sum_{m,m^{\prime},m_S,m_L}C^{S,m_S}_{1m,1m^{\prime}}C^{J,M}_{Sm_S,Lm_L}\epsilon_{m}^\mu\epsilon_{m^{\prime}}^\nu Y_{L,m_L},\label{spin-orbitalwavefunctions2}
\end{eqnarray}
respectively. In the above expressions, the symbols $C^{e,f}_{ab,cd}$ and $Y_{L,m_L}$ are used to denote the Clebsch-Gordan coefficient and the spherical harmonics function, respectively.

By utilising the established principal steps of the OBE model \cite{Chen:2016qju}, we can derive the effective interactions in the coordinate space of the $\bar K^{(*)} \bar K^{(*)}\to \bar K^{(*)} \bar K^{(*)}$ processes, which are given by
\begin{eqnarray}
{V}_{I,J}^{\bar K \bar K \to \bar K \bar K}&=&-g^{\prime 2}_{s}Y_{\sigma}+\frac{1}{2}\beta^{\prime 2}g^2_V\mathcal{G}_{\mathbb{V}}(I)Y_{\mathbb{V}},\label{effectiveinteractions0}\\
{V}_{I,J}^{\bar K \bar K^* \to \bar K \bar K^*}&=&-g^{\prime 2}_{s}\mathcal{O}_{1}[J]Y_{\sigma}+\frac{1}{2}\beta^{\prime 2}g^2_V\mathcal{O}_{1}[J]\mathcal{G}_{\mathbb{V}}(I)Y_{\mathbb{V}},\label{effectiveinteractions1}\\
{V}_{I,J}^{\bar K \bar K^* \to \bar K^* \bar K}&=&\frac{g^{\prime 2}}{3f^2_{\pi}}\left(\mathcal{O}_{2}[J]\mathcal{Z}_{r}+\mathcal{O}_{3}[J]\mathcal{T}_{r}\right)\mathcal{H}_{\mathbb{P}}^{\prime}(I)Y_{\mathbb{P}0}\nonumber\\
&+&\frac{2}{3}\lambda^{\prime 2}g^2_V\left(2\mathcal{O}_{2}[J]\mathcal{Z}_{r}-\mathcal{O}_{3}[J]\mathcal{T}_{r}\right)\mathcal{G}_{\mathbb{V}}^{\prime}(I)Y_{\mathbb{V}0},\nonumber\\\\
{V}_{I,J}^{\bar K^* \bar K^* \to \bar K^* \bar K^*}&=&-g^{\prime 2}_{s}\mathcal{O}_4[J]Y_{\sigma}\nonumber\\
&-&\frac{g^{\prime 2}}{3f^2_{\pi}}\left(\mathcal{O}_{5}[J]\mathcal{Z}_{r}+\mathcal{O}_{6}[J]\mathcal{T}_{r}\right)\mathcal{H}_{\mathbb{P}}(I)Y_{\mathbb{P}}\nonumber\\
&+&\frac{1}{2}\beta^{\prime 2}g^2_V\mathcal{O}_4[J]\mathcal{G}_{\mathbb{V}}(I)Y_{\mathbb{V}}\nonumber\\
&-&\frac{2}{3}\lambda^{\prime 2}g^2_V\left(2\mathcal{O}_{5}[J]\mathcal{Z}_{r}-\mathcal{O}_{6}[J]\mathcal{T}_{r}\right)\mathcal{G}_{\mathbb{V}}(I)Y_{\mathbb{V}},\nonumber\\\\
{V}_{I,J}^{\bar K \bar K \to \bar K^* \bar K^*}&=&\frac{g^{\prime 2}}{3f^2_{\pi}}\left(\mathcal{O}_{7}[J]\mathcal{Z}_{r}+\mathcal{O}_{8}[J]\mathcal{T}_{r}\right)\mathcal{H}_{\mathbb{P}}(I)Y_{\mathbb{P}}\nonumber\\
&+&\frac{2}{3}\lambda^{\prime 2}g^2_V\left(2\mathcal{O}_{7}[J]\mathcal{Z}_{r}-\mathcal{O}_{8}[J]\mathcal{T}_{r}\right)\mathcal{G}_{\mathbb{V}}(I)Y_{\mathbb{V}},\nonumber\\\\
{V}_{I,J}^{\bar K \bar K^* \to \bar K^* \bar K^*}&=&\frac{g^{\prime 2}}{3f^2_{\pi}}\left(\mathcal{O}_{9}[J]\mathcal{Z}_{r}+\mathcal{O}_{10}[J]\mathcal{T}_{r}\right)\mathcal{H}_{\mathbb{P}}^{\prime\prime}(I)Y_{\mathbb{P}1}\nonumber\\
&+&\frac{2}{3}\lambda^{\prime 2}g^2_V\left(2\mathcal{O}_{9}[J]\mathcal{Z}_{r}-\mathcal{O}_{10}[J]\mathcal{T}_{r}\right)\mathcal{G}_{\mathbb{V}}^{\prime\prime}(I)Y_{\mathbb{V}1},\nonumber\\\\
{V}_{I,J}^{\bar K^* \bar K \to \bar K^* \bar K^*}&=&-\frac{g^{\prime 2}}{3f^2_{\pi}}\left(\mathcal{O}_{11}[J]\mathcal{Z}_{r}+\mathcal{O}_{12}[J]\mathcal{T}_{r}\right)\mathcal{H}_{\mathbb{P}}^{\prime\prime}(I)Y_{\mathbb{P}1}\nonumber\\
&-&\frac{2}{3}\lambda^{\prime 2}g^2_V\left(2\mathcal{O}_{11}[J]\mathcal{Z}_{r}-\mathcal{O}_{12}[J]\mathcal{T}_{r}\right)\mathcal{G}_{\mathbb{V}}^{\prime\prime}(I)Y_{\mathbb{V}1}.\label{effectiveinteractions2}\nonumber\\
\end{eqnarray}
In Eqs.~(\ref{effectiveinteractions0})-(\ref{effectiveinteractions2}), the function $Y_i$ is defined as follows:
\begin{eqnarray}
Y_i = \left\{
\begin{aligned}
|q_i|&\leqslant m,\ \frac{e^{-m_i r}-e^{-\Lambda^2_i r}}{4\pi r}-\frac{\Lambda^2_i-m^2_i}{8\pi\Lambda_i}e^{-\Lambda_i r},\\
|q_i|&>m,\ \frac{\mathrm{cos} (m^{\prime}_i r)-e^{-\Lambda_i r}}{4\pi r}-\frac{\Lambda^2_i+m^{\prime2}_i}{8\pi\Lambda_i}e^{-\Lambda_i r},
\end{aligned}
\right.
\end{eqnarray}
where the following definitions are applicable:
\begin{eqnarray}
m_i=\sqrt{m^2-q^2_i},~~~~m^{\prime}_i=\sqrt{q^2_i-m^2},~~{\rm and}~~\Lambda_i=\sqrt{\Lambda^2-q^2_i}\nonumber\\
\end{eqnarray}
with $q_0=0.396\,{\rm GeV}$ and $q_1=0.154\,{\rm GeV}$. The operators $\mathcal{Z}_{r}$ and $\mathcal{T}_{r}$ act on the function $Y_i$ are defined as
\begin{eqnarray}
\mathcal{Z}_{r}=\frac{1}{r^2}\frac{\partial}{\partial r}r^2\frac{\partial}{\partial r}~~~~~{\rm and}~~~~~
\mathcal{T}_{r}=r\frac{\partial}{\partial r}\frac{1}{r}\frac{\partial}{\partial r}.
\end{eqnarray}
The isospin factors, designated as $\mathcal{H}_{\mathbb{P}}^{(\prime\,,\prime\prime)}$ and $\mathcal{G}_{\mathbb{V}}^{(\prime\,,\prime\prime)}$, associated with the flavor wave functions of the $\bar K^{(*)} \bar K^{(*)}$ systems listed in Eqs.~(\ref{flavorwavefunctions1})-(\ref{flavorwavefunctions2}) are defined as follows \cite{Li:2012ss}:
\begin{eqnarray}
\renewcommand\tabcolsep{2.10cm}
\renewcommand{\arraystretch}{1.50}
\begin{array}{*{3}ccc}
\hline
&&\mathcal{H}_{\pi}(0)=-\frac{3}{2}&~~~~~~~\mathcal{H}_{\eta}(0)=\frac{1}{6}&\nonumber\\
&&\mathcal{G}_{\rho}(0)=-\frac{3}{2}&~~~~~~~\mathcal{G}_{\omega}(0)=\frac{1}{2}&~~~~~~~\mathcal{G}_{\phi}(0)=1\nonumber\\
&&\mathcal{H}_{\pi}(1)=\frac{1}{2}&~~~~~~~\mathcal{H}_{\eta}(1)=\frac{1}{6}&\nonumber\\
&&\mathcal{G}_{\rho}(1)=\frac{1}{2}&~~~~~~~\mathcal{G}_{\omega}(1)=\frac{1}{2}&~~~~~~~\mathcal{G}_{\phi}(1)=1\nonumber\\
\hline
&&\mathcal{H}_{\pi}^{\prime}(0)=\frac{3}{2}&~~~~~~~\mathcal{H}_{\eta}^{\prime}(0)=-\frac{1}{6}&\nonumber\\
&&\mathcal{G}_{\rho}^{\prime}(0)=\frac{3}{2}&~~~~~~~\mathcal{G}_{\omega}^{\prime}(0)=-\frac{1}{2}&~~~~~~~\mathcal{G}_{\phi}^{\prime}(0)=-1\nonumber\\
&&\mathcal{H}_{\pi}^{\prime}(1)=\frac{1}{2}&~~~~~~~\mathcal{H}_{\eta}^{\prime}(1)=\frac{1}{6}&\nonumber\\
&&\mathcal{G}_{\rho}^{\prime}(1)=\frac{1}{2}&~~~~~~~\mathcal{G}_{\omega}^{\prime}(1)=\frac{1}{2}&~~~~~~~\mathcal{G}_{\phi}^{\prime}(1)=1\nonumber\\
\hline
&&\mathcal{H}^{\prime\prime}_{\pi}(0)=-\frac{3}{2\sqrt{2}}&~~~~~~~\mathcal{H}^{\prime\prime}_{\eta}(0)=\frac{1}{6\sqrt{2}}&\nonumber\\
&&\mathcal{G}^{\prime\prime}_{\rho}(0)=-\frac{3}{2\sqrt{2}}&~~~~~~~\mathcal{G}^{\prime\prime}_{\omega}(0)=\frac{1}{2\sqrt{2}}&~~~~~~~\mathcal{G}^{\prime\prime}_{\phi}(0)=\frac{1}{\sqrt{2}}\\
\hline
\end{array}.
\end{eqnarray}
Furthermore, the spin-dependent operators in Eqs.~(\ref{effectiveinteractions1})-(\ref{effectiveinteractions2}), designated as $\mathcal{O}_{i}[J]$, are defined as follows:
\begin{eqnarray}
\mathcal{O}_{1}[J]&=&{\bm\epsilon^{\dagger}_4}\cdot{\bm\epsilon_2},\nonumber\\
\mathcal{O}_{2}[J]&=&{\bm\epsilon^{\dagger}_3}\cdot{\bm\epsilon_2},~~~~~~~~~~~~~~~~~~\mathcal{O}_{3}[J]=S({\bm\epsilon^{\dagger}_3},{\bm\epsilon_2},\hat{\bm{r}}),\nonumber\\
\mathcal{O}_{4}[J]&=&\left({\bm\epsilon^{\dagger}_3}\cdot{\bm\epsilon_1}\right)\left({\bm\epsilon^{\dagger}_4}\cdot{\bm\epsilon_2}\right),~\mathcal{O}_{5}[J]=\left({\bm\epsilon^{\dagger}_3}\times{\bm\epsilon_1}\right)\cdot\left({\bm\epsilon^{\dagger}_4}\times{\bm\epsilon_2}\right),\nonumber\\
\mathcal{O}_{6}[J]&=&S({\bm\epsilon^{\dagger}_3}\times{\bm\epsilon_1},{\bm\epsilon^{\dagger}_4}\times{\bm\epsilon_2},\hat{\bm{r}}),\nonumber\\
\mathcal{O}_{7}[J]&=&{\bm\epsilon^{\dagger}_4}\cdot{\bm\epsilon^{\dagger}_3},~~~~~~~~~~~~~~~~~~
\mathcal{O}_{8}[J]=S({\bm\epsilon^{\dagger}_4},{\bm\epsilon^{\dagger}_3},\hat{\bm{r}}),\nonumber\\ \mathcal{O}_{9}[J]&=&{\bm\epsilon^{\dagger}_3}\cdot\left({\bm\epsilon^{\dagger}_4}\times{\bm\epsilon_2}\right),~~~~~
\mathcal{O}_{10}[J]=S(\bm\epsilon^{\dagger}_3,{\bm\epsilon^{\dagger}_4}\times{\bm\epsilon_2},\hat{\bm{r}}),\nonumber\\
\mathcal{O}_{11}[J]&=&{\bm\epsilon^{\dagger}_4}\cdot\left({\bm\epsilon^{\dagger}_3}\times{\bm\epsilon_1}\right),~~~~~
\mathcal{O}_{12}[J]=S(\bm\epsilon^{\dagger}_4,{\bm\epsilon^{\dagger}_3}\times{\bm\epsilon_1},\hat{\bm{r}}),
\end{eqnarray}
where
\begin{eqnarray}
S({\bm a},{\bm b},\hat{\bm{r}})= 3\left(\hat{\bm r} \cdot {\bm a}\right)\left(\hat{\bm r} \cdot {\bm b}\right)-{\bm a} \cdot {\bm b}
\end{eqnarray}
is the tensor force operator. In the concrete calculations, we can deduce the numerical matrix elements $\langle f|\mathcal{O}_i[J]|i\rangle$ by utilising the operators $\mathcal{O}_i[J]$ sandwiched between the spin-orbit wave functions $|^{2S+1}L_J\rangle$ of the $\bar K^{(*)} \bar K^{(*)}$ systems listed in Eqs.~(\ref{spin-orbitalwavefunctions1})-(\ref{spin-orbitalwavefunctions2}). To illustrate, the following numerical matrix elements can be obtained
\begin{eqnarray}
&&\mathcal{O}_1[1] \mapsto \rm {diag}(1,1),~~~~~~~~~~~~~~~\mathcal{O}_2[1] \mapsto \rm {diag}(1,1),\nonumber\\
&&\mathcal{O}_{3}[1] \mapsto \left(\begin{array}{cc} 0 & -\sqrt{2} \\ -\sqrt{2} & 1\end{array}\right),\nonumber\\
&&\mathcal{O}_4[0] \mapsto \rm {diag}(1,1),~~~~~~~~~~~~~~~ \mathcal{O}_5[0] \mapsto \rm {diag}(2,-1),\nonumber\\
&&\mathcal{O}_{6}[0] \mapsto \left(\begin{array}{cc} 0 & \sqrt{2} \\ \sqrt{2} & 2\end{array}\right),\nonumber\\
&&\mathcal{O}_4[1] \mapsto \rm {diag}(1,1),~~~~~~~~~~~~~~~\mathcal{O}_5[1] \mapsto \rm {diag}(1,1),\nonumber\\
&&\mathcal{O}_{6}[1] \mapsto \left(\begin{array}{cc} 0 & -\sqrt{2} \\ -\sqrt{2} & 1\end{array}\right),\nonumber\\
&&\mathcal{O}_4[2] \mapsto \rm {diag}(1,1,1),~~~~~~~~~~~ \mathcal{O}_5[2] \mapsto \rm {diag}(-1,2,-1),\nonumber\\
&&\mathcal{O}_{6}[2] \mapsto \left(\begin{array}{ccc} 0 & \frac{\sqrt{2}}{\sqrt{5}} &-\frac{\sqrt{14}}{\sqrt{5}} \\ \frac{\sqrt{2}}{\sqrt{5}} & 0 &-\frac{2}{\sqrt{7}}\\ -\frac{\sqrt{14}}{\sqrt{5}}&-\frac{2}{\sqrt{7}}&-\frac{3}{7}\end{array}\right),
\end{eqnarray}
when considering $S$-$D$ wave mixing effect for the $\bar K^{(*)} \bar K^{(*)}$ systems.

\section{Spectroscopic properties of the double-strangeness molecular tetraquarks}\label{sec3}

This section presents the bound-state properties of the $\bar K^{(*)} \bar K^{(*)}$ systems, along with an analysis of the strong decay behaviors, magnetic moments, and M1 radiative decay widths of the $\bar K^{(*)} \bar K^{(*)}$ molecular tetraquark candidates.

\subsection{Bound-state properties of the $\bar K^{(*)} \bar K^{(*)}$ systems}

The effective interactions of the $\bar K^{(*)} \bar K^{(*)}$ systems listed in Eqs.~(\ref{effectiveinteractions0})-(\ref{effectiveinteractions2}) allow for further discussion of their bound-state properties by solving the coupled-channel Schr$\ddot{\rm o}$dinger equation, which  will facilitate the prediction of the mass spectra of double-strangeness molecular tetraquarks.

In the concrete calculation, it is necessary to determine the coupling constants associated with the effective interactions of the $\bar K^{(*)} \bar K^{(*)}$ systems, which are of particular importance when attempting to search for loosely bound-state solutions of the $\bar K^{(*)} \bar K^{(*)}$ systems. In principle, it is our preference to fix the coupling constants by fitting the experimental information, when the relevant experimental data exists. For example, the value of $g^{\prime}$ can be ascertained from the decay width of the $K^* \to K \pi$ process \cite{ParticleDataGroup:2022pth}, which yields $g^{\prime}=1.12$ \cite{Wang:2024ukc}. In the absence of pertinent experimental data, the coupling constants can be deduced on the basis of the theoretical models and approaches. In this instance, the quark model \cite{Riska:2000gd} enables the determination of the values of $g_{s}^{\prime}=0.76$ and $\lambda^{\prime}=0.56$ GeV$^{-1}$ \cite{Liu:2020nil,Wang:2024ukc}, while the hidden-gauge symmetry of the vector meson gives rise to the result of $\beta^{\prime}=0.835$ \cite{Molina:2010tx}. Furthermore, the values of $f_{\pi}=0.132$ GeV and $g_V=5.8$ \cite{Bando:1987br,Isola:2003fh} are adopted in this study. In addition, the meson masses are required, which include
$m_{\sigma}=600.00~\rm{MeV}$, $m_{\pi}=137.27~\rm{MeV}$,
$m_{\eta} =547.86~\rm{MeV}$, $m_{\rho}=775.26~\rm{MeV}$,
$m_{\omega}=782.66~\rm{MeV}$, $m_{\phi}=1019.46~\rm{MeV}$,
$m_K=495.65~\rm{MeV}$, and $m_{K^{\ast}}=891.67~\rm{MeV}$ \cite{ParticleDataGroup:2022pth}.

Firstly, we focus on the effective interaction of the $\bar K \bar K$ system and the resulting formation of double-strangeness molecular tetraquark candidate. As a consequence of the symmetry constraints, the effective interaction of the $I(J^P)=1(0^+)$ $\bar K \bar K$ state is relatively simple, with only the exchange of four light bosons, namely the $\sigma$, $\rho$, $\omega$, and $\phi$ mesons, being possible (see Eq.~(\ref{effectiveinteractions0})). Among these, the effective interactions associated with the $\rho$, $\omega$, and $\phi$ exchanges are repulsive, while the $\sigma$ exchange interaction exhibits a slight attractive character at the intermediate range. It can thus be concluded that the total effective interaction of the $I(J^P)=1(0^+)$ $\bar K \bar K$ state is repulsive. Following the solutions of the coupled-channel Schr$\ddot{\rm o}$dinger equation, the single-channel and coupled-channel analysis indicates that the $I(J^P)=1(0^+)$ $\bar K \bar K$ state does not exhibit loosely bound-state solutions until the cutoff value is increased to approximately 2 GeV. Thus, the $I(J^P)=1(0^+)$ $\bar K \bar K$ state is unlikely to be double-strangeness molecular tetraquark candidate.

\begin{figure}[htbp]
  \centering
  \includegraphics[width=6.0cm]{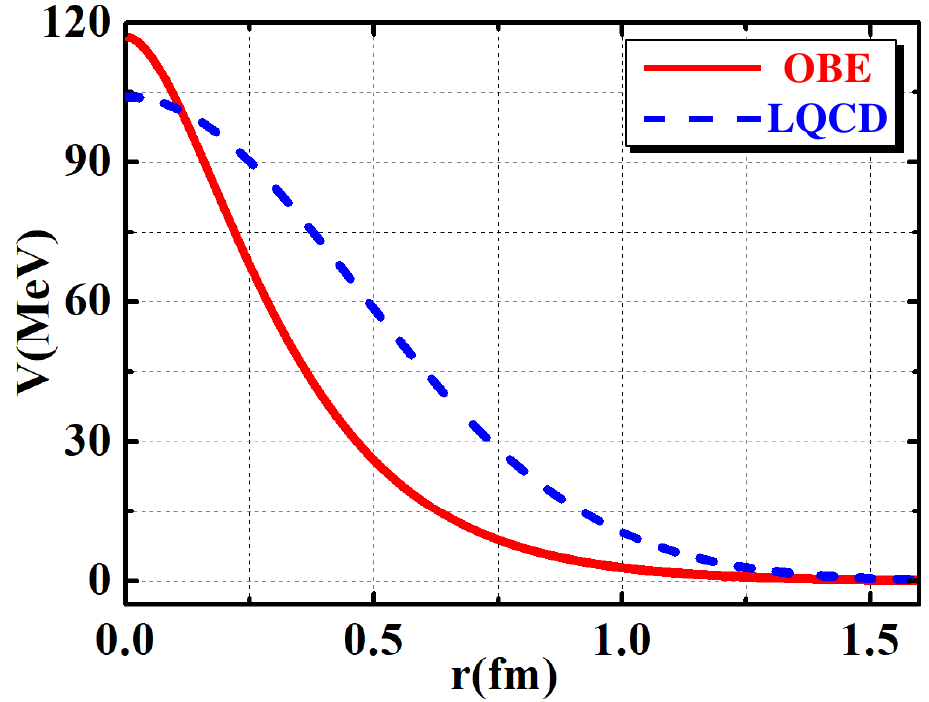}
  \caption{The effective interaction of the $I(J^P)=1(0^+)$ $\bar K \bar K$ state as predicted by the OBE model and the lattice QCD calculation \cite{Kanada-Enyo:2008wsu}, which are shown by the red solid and blue dashed lines, respectively. In the OBE effective interaction, the cutoff value is fixed at 1.3 GeV.}\label{OBEandLattice}
\end{figure}

In Ref. \cite{Beane:2007uh}, the lattice QCD calculation yielded the $I(J^P)=1(0^+)$ $K^+K^+$ scattering length as $a_{[K^+K^+][1(0^+)]}=-0.141\pm0.006\,{\rm fm}$, indicating that the effective interaction of the $I(J^P)=1(0^+)$ $\bar K \bar K$ state is repulsive. It can thus be concluded that the OBE effective interaction of the $I(J^P)=1(0^+)$ $\bar K \bar K$ state is in qualitative agreement with the result obtained from the lattice QCD calculation \cite{Beane:2007uh}. Furthermore, the authors proposed that the effective interaction of the $I(J^P)=1(0^+)$ $\bar K \bar K$ state can be approximated by a Gaussian function in Ref. \cite{Kanada-Enyo:2008wsu}, which is given by
\begin{eqnarray}
{V}_{[\bar K \bar K][1(0^+)]}=U_{[\bar K \bar K]}^{I=1}e^{-(r/b)^2},
\end{eqnarray}
where the interaction strength, $U_{[\bar K \bar K]}^{I=1}$, and the interaction range, $b$, are the parameters. By reproducing the lattice QCD scattering length of the $I(J^P)=1(0^+)$ $K^+K^+$ state \cite{Beane:2007uh}, the values of $U_{[\bar K \bar K]}^{I=1}$ and $b$ can be determined to be $104\,{\rm MeV}$ and $0.66\,{\rm fm}$ \cite{Kanada-Enyo:2008wsu}, respectively. In Fig. \ref{OBEandLattice}, we present the effective interaction of the $I(J^P)=1(0^+)$ $\bar K \bar K$ state as predicted by the OBE model and the lattice QCD calculation \cite{Kanada-Enyo:2008wsu}. As illustrated in Fig. \ref{OBEandLattice}, it can be concluded that the OBE effective interaction of the $I(J^P)=1(0^+)$ $\bar K\bar K$ state is in accordance with the result of the lattice QCD calculation \cite{Kanada-Enyo:2008wsu}, exhibiting a quantitative agreement.

Subsequently, we discuss the effective interactions of the $\bar K \bar K^*$ and $\bar K^* \bar K^*$ systems, and the resulting formation of double-strangeness molecular tetraquark candidates. Until the cutoff value $\Lambda$ is increased to approximately 2 GeV, it is not possible to identify loosely bound-state solutions for the $I=1$ $\bar K \bar K^*$ and $\bar K^* \bar K^*$ systems. This remains the case even when the contribution of $S$-$D$ wave mixing and coupled-channel effects is included. It can thus be concluded that the results of our analysis do not support the existence of double-strangeness molecular tetraquark candidates for the $I=1$ $\bar K \bar K^*$ and $\bar K^* \bar K^*$ systems, which exhibits the similar bound-state properties for the $I=1$ $D^{(*)} D^{(*)}$ systems \cite{Li:2012ss}.

\begin{figure}[htbp]
  \centering
  \includegraphics[width=8.8cm]{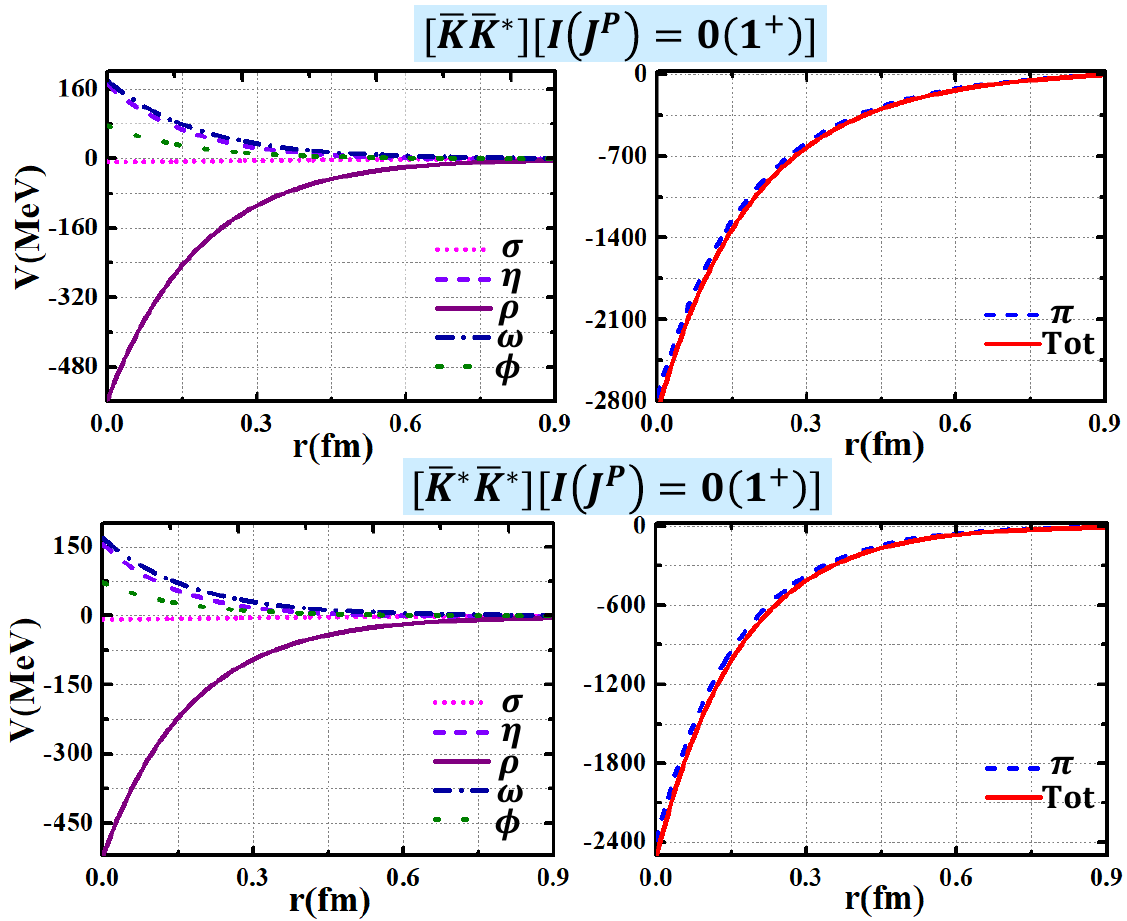}
  \caption{The OBE effective interactions of the $I(J^P)=0(1^+)$ $\bar K \bar K^*$ and  $I(J^P)=0(1^+)$ $\bar K^* \bar K^*$ states, where the cutoff $\Lambda$ is fixed at a value of 1.2 GeV.}\label{KKstarKstarKstarpotential}
\end{figure}

In the following, we primarily focus on the effective interactions and bound-state properties of the $I(J^P)=0(1^+)$ $\bar K \bar K^*$ and $I(J^P)=0(1^+)$ $\bar K^* \bar K^*$ states. Fig. \ref{KKstarKstarKstarpotential} depicts the effective interactions of the $I(J^P)=0(1^+)$ $\bar K \bar K^*$ and $I(J^P)=0(1^+)$ $\bar K^* \bar K^*$ states. As shown in Fig. \ref{KKstarKstarKstarpotential}, the effective interaction of the $I(J^P)=0(1^+)$ $\bar K \bar K^*$ state is analogous to that of the $I(J^P)=0(1^+)$ $\bar K^* \bar K^*$ state. Furthermore, it is apparent that: (1) The $\pi$ exchange interaction provides a dominant contribution to the effective interactions of the $I(J^P)=0(1^+)$ $\bar K \bar K^*$ and $I(J^P)=0(1^+)$ $\bar K^* \bar K^*$ states, which is significantly greater than other effective interactions. This finding aligns with the conclusion that the $\pi$ exchange interaction plays an indispensable role in the generation of loosely bound state within the OBE model \cite{Chen:2016qju}. Here, it is imperative to underscore that the interaction strength of the $\pi$ exchange has been determined from the experimental width of the $K^* \to K \pi$ process \cite{ParticleDataGroup:2022pth}, which can serve as a reliable estimate of the $\pi$ exchange interaction. (2) In regard to the vector meson exchange interactions, the effective interaction derived from the $\rho$ exchange is characterised by strongly attractive, whereas the $\omega$ and $\phi$ exchange effective interactions are repulsive, which aligns with the conclusions presented in Ref. \cite{Chen:2017vai}. In quantitative terms, the interaction from the $\rho$ exchange is approximately three times stronger than that provided by the $\omega$ exchange, and the contribution of the $\phi$ exchange effective interaction can be considered negligible in comparison to that of the $\rho$ exchange effective interaction.

In Table \ref{KKstarboundstate}, we present the bound-state properties of the $I(J^P)=0(1^+)$ $\bar K \bar K^*$ state by considering the single-channel, $S$-$D$ wave mixing, and coupled-channel analysis. For the $I(J^P)=0(1^+)$ $\bar K \bar K^*$ state, there exists loosely bound-state solutions when the cutoff value is tuned to approximately 1.21 GeV by the single-channel analysis. Following the incorporation of the contribution of the $D$-wave channel, loosely bound-state solutions can be obtained when the cutoff value is lowered to 1.03 GeV, and this loosely bound state is predominantly composed of the $[\bar K \bar K^*]|{}^3\mathbb{S}_{1}\rangle$ channel. After including the contribution of coupled-channel effect, our numerical findings indicate that loosely bound-state solutions can be obtained by selecting a cutoff value of approximately 0.79 GeV or even larger, and the $[\bar K \bar K^*]|{}^3\mathbb{S}_{1}\rangle$ channel is the dominant channel. Incorporating the influence of $S$-$D$ wave mixing and coupled-channel effects, the cutoff value becomes smaller when obtaining the same binding energy. Consequently, $S$-$D$ wave mixing and coupled-channel effects exert a beneficial influence on the formation of the $I(J^P)=0(1^+)$ $\bar K \bar K^*$ bound state. In general, a bound state with the small binding energy and the large root-mean-square radius is the most likely candidate for hadronic molecular state when taking a cutoff value around 1 GeV \cite{Chen:2016qju}. This evaluation leads to the conclusion that the $I(J^P)=0(1^+)$ $\bar K \bar K^*$ state is the most likely candidate for double-strangeness molecular tetraquark, which is consistent with the conclusion based on the chiral quark model \cite{Ji:2024znj}.

\renewcommand\tabcolsep{0.03cm}
\renewcommand{\arraystretch}{1.50}
\begin{table}[!htbp]
\centering
\caption{Bound-state properties of the $I(J^P)=0(1^+)$ $\bar K \bar K^*$ state by considering three analytical scenarios: (I) the single-channel analysis, (II) the $S$-$D$ wave mixing analysis, and (III) the coupled-channel analysis. Here, the symbols $P_1$, $P_2$, $P_3$, and $P_4$ represent the probabilities of the $[\bar K \bar K^*]|{}^3\mathbb{S}_{1}\rangle$, $[\bar K \bar K^*]|{}^3\mathbb{D}_{1}\rangle$, $[\bar K^* \bar K^*]|{}^3\mathbb{S}_{1}\rangle$, and $[\bar K^* \bar K^*]|{}^3\mathbb{D}_{1}\rangle$ channels, respectively.}\label{KKstarboundstate}
\begin{tabular}{c|ccccccc}\toprule[1.0pt]\toprule[1.0pt]
Scenarios&$\Lambda\,(\rm{GeV})$  &$E\,(\rm {MeV})$ &$r_{\rm RMS}\,(\rm {fm})$ &$P_1\,(\%)$&$P_2\,(\%)$&$P_3\,(\%)$&$P_4\,(\%)$\\\midrule[1.0pt]
\multirow{3}{*}{I}&1.21&$-4.27$&2.83&100.00\\
&1.27&$-14.83$&1.47&100.00\\
&1.33&$-30.02$&1.05&100.00\\\midrule[1.0pt]
\multirow{3}{*}{II}&1.03&$-3.88$&3.86&95.98&4.02     \\
&1.11&$-13.01$ &1.90&96.76&3.24    \\
&1.19&$-30.37$&1.19&96.60&3.40     \\\midrule[1.0pt]
\multirow{3}{*}{III}&0.79&$-3.86$&3.88&94.08&3.85&2.01&0.06     \\
&0.83&$-12.50$&1.91&91.73&1.93&6.24&0.10\\
&0.87&$-30.49$&1.14&88.13&1.32&10.44&0.11    \\
\bottomrule[1.0pt]\bottomrule[1.0pt]
\end{tabular}
\end{table}

In Table \ref{KstarKstarboundstate}, we collect the bound-state properties of the $I(J^P)=0(1^+)$ $\bar K^* \bar K^*$ state by considering both the single-channel and $S$-$D$ wave mixing analysis. For the $I(J^P)=0(1^+)$ $\bar K^* \bar K^*$ state, the bound-state solution with a relatively small binding energy and an appropriate root-mean-square radius can be obtained by performing the single-channel analysis, when $\Lambda$ is set to a value slightly greater than 1.23 GeV. After incorporating $S$-$D$ wave mixing effect into our calculation, we find that loosely bound-state solutions can be obtained when we set the cutoff value $\Lambda$ around 1.11 GeV, and the $|{}^3\mathbb{S}_{1}\rangle$ channel plays a major role in the formation of the $I(J^P)=0(1^+)$ $\bar K^* \bar K^*$ bound state. A comparison of the results with and without $S$-$D$ wave mixing effect demonstrates that the properties of loosely bound states are modified in accordance with the inclusion of the contribution of the $D$-wave channel. Given that the $I(J^P)=0(1^+)$ $\bar K^* \bar K^*$ bound state exhibits the shallow binding energy and the suitable root-mean-square radius within the reasonable range of the cutoff value \cite{Chen:2016qju}, we conclude that the $I(J^P)=0(1^+)$ $\bar K^* \bar K^*$ state is the most likely candidate for double-strangeness molecular tetraquark.

\renewcommand\tabcolsep{0.12cm}
\renewcommand{\arraystretch}{1.50}
\begin{table}[!htbp]
\centering
\caption{Bound-state properties of the $I(J^P)=0(1^+)$ $\bar K^* \bar K^*$ state by considering two analytical scenarios: (I) the single-channel analysis and (II) the $S$-$D$ wave mixing analysis.}\label{KstarKstarboundstate}
\begin{tabular}{c|ccccc}\toprule[1.0pt]\toprule[1.0pt]
Scenarios&\multicolumn{5}{c}{Bound-state properties}\\\hline
\multirow{4}{*}{I}&$\Lambda\,(\rm{GeV})$  &$E\,(\rm {MeV})$ &$r_{\rm RMS}\,(\rm {fm})$\\\cline{2-6}
&1.23&$-4.55$&2.31 \\
&1.30&$-15.28$&1.34\\
&1.36&$-29.95$&1.00\\\midrule[1.0pt]
\multirow{4}{*}{II}&$\Lambda\,(\rm{GeV})$  &$E\,(\rm {MeV})$ &$r_{\rm RMS}\,(\rm {fm})$&$P_{\left|{}^3\mathbb{S}_{1}\right\rangle}\,(\%)$&$P_{\left|{}^3\mathbb{D}_{1}\right\rangle}\,(\%)$\\\cline{2-6}
&1.11&$-3.99$&2.56&98.02&1.98\\
&1.19&$-14.45$&1.46&97.43&2.57\\
&1.27&$-32.64$&1.04&97.18&2.82\\
\bottomrule[1.0pt]\bottomrule[1.0pt]
\end{tabular}
\end{table}

In the context of discussing the bound-state properties of the systems by solving the Schr\"{o}dinger equation $H \psi(r)=E \psi(r)$, the Hamiltonian $H$ contains the kinetic energy term $T$ and the  potential energy term $V$, i.e., $H=T+V$. Among them, the kinetic energy term is related to the reduced mass of the system $\mu_{[H_1H_2]}=m_{H_1}m_{H_2}/(m_{H_1}+m_{H_2})$. Given the width of the $K^*$ meson, it is necessary to consider the mass distribution of the $K^*$ meson when determining the reduced mass of the $\bar K^* \bar K^*$ system. In the following, we take three typical values $841.67, 891.67$, and $941.67~\rm{MeV}$ for the mass distribution of the $K^*$ meson when determining the reduced mass of the $\bar K^* \bar K^*$ system, which correspond to the lower, central, and upper values of the mass distribution of the $K^*$ meson, respectively\footnote{We would like to thank the referee for suggesting that we discuss the bound-state properties by considering the mass distribution of the $K^*$ (with a width of approximately 50 MeV), which is a component of the molecular systems under discussion.}. Table~\ref{sr} displays the obtained bound-state solutions for the $I(J^P)=0(1^+)$ $\bar K^* \bar K^*$ state, with twice the reduced mass of the $\bar K^* \bar K^*$ system sets to $2\mu_{[\bar K^* \bar K^*]}=841.67, 891.67$, and $941.67~\rm{MeV}$. In light of the mass distribution of the $K^*$ meson, the numerical results pertaining to the bound state properties of the $I(J^P)=0(1^+)$ $\bar K^* \bar K^*$ state may change to some extent, but the conclusion of the $I(J^P)=0(1^+)$ $\bar K^* \bar K^*$ state as the most likely candidate for double-strangeness molecular tetraquark remains unchanged. In particular, when $2\mu_{[\bar K^* \bar K^*]}=891.67~\rm{MeV}$ and $\Lambda=1.23~\rm{GeV}$, the binding energy of the $I(J^P)=0(1^+)$ $\bar K^* \bar K^*$ state is found to be $-4.55~\rm{MeV}$. If $2\mu_{[\bar K^* \bar K^*]}=841.67~\rm{MeV}$ and $\Lambda=1.23~\rm{GeV}$, it is found that the $I(J^P)=0(1^+)$ $\bar K^* \bar K^*$ state still has the loosely bound state solution, with a binding energy of $-1.10~\rm{MeV}$.

\renewcommand\tabcolsep{0.22cm}
\renewcommand{\arraystretch}{1.50}
\begin{table}[!htbp]
\centering
\caption{Bound-state solutions for the $I(J^P)=0(1^+)$ $\bar K^* \bar K^*$ state by taking $2\mu_{[\bar K^* \bar K^*]}=841.67, 891.67$, and $941.67~\rm{MeV}$. The units of the cutoff $\Lambda$, binding energy $E$, and root-mean-square radius $r_{{\rm RMS}}$ are $\rm{GeV}$, $\rm {MeV}$, and $\rm {fm}$, respectively.}\label{sr}
\begin{tabular}{c|cc|cc|cc}\toprule[1pt]\toprule[1pt]
$2\mu_{[K^{*}K^{*}]}$&\multicolumn{2}{c|}{$841.67~\rm{MeV}$}&\multicolumn{2}{c|}{$891.67~\rm{MeV}$}&\multicolumn{2}{c}{$941.67~\rm{MeV}$}\\\midrule[1.0pt]
$\Lambda$ &$E$&$r_{\rm RMS}$        &$E$&$r_{\rm RMS}$           &$E$&$r_{\rm RMS}$\\
1.23&$-1.10$ &4.22                  &$-4.55$&2.31                   &$-9.60$&1.61\\
1.30&$-7.96$ &1.83                  &$-15.28$&1.34                     &$-24.05$&1.08\\
1.36&$-18.92$ &1.25                 &$-29.95$&1.00                    &$-42.32$&0.85\\
\bottomrule[1pt]\bottomrule[1pt]
\end{tabular}
\end{table}

In conclusion, the $I(J^P)=0(1^+)$ $\bar K \bar K^*$  and $I(J^P)=0(1^+)$ $\bar K^* \bar K^*$ states are the most likely candidates for double-strangeness molecular tetraquarks. Thus, we propose that the future experiments should concentrate the $I(J^P)=0(1^+)$ $\bar K \bar K^*$ and $I(J^P)=0(1^+)$ $\bar K^* \bar K^*$ molecular states by the weak decays of the $B$ meson, such as LHCb, Belle II, and other relevant experiments.

\subsection{Strong decay behaviors of the $I(J^P)=0(1^+)$ $\bar K \bar K^*$ and $I(J^P)=0(1^+)$ $\bar K^* \bar K^*$ molecular candidates}

In order to provide valuable information for the experimental search for the $I(J^P)=0(1^+)$ $\bar K \bar K^*$ and $I(J^P)=0(1^+)$ $\bar K^* \bar K^*$ molecular candidates, we further estimate their typical strong decay behaviors in this subsection.

\begin{figure}[htbp]
    \centering
    \includegraphics[width=8.7cm]{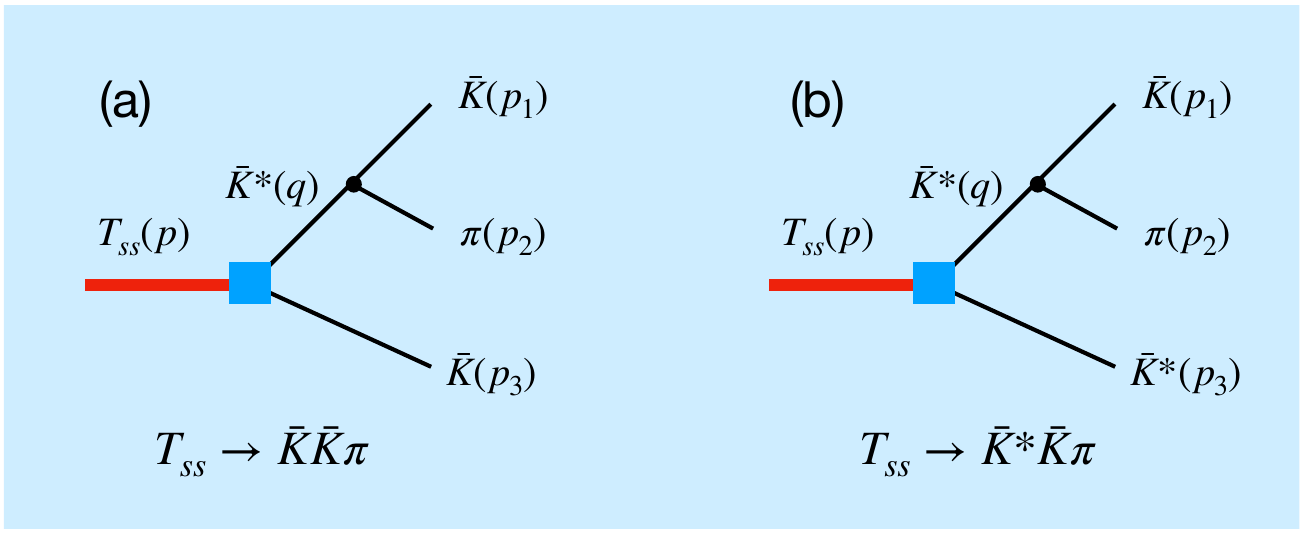}
    \caption{The diagrams for depicting the $[\bar K \bar K^*][0(1^+)] \to \bar K \bar K \pi$ and $[\bar K^* \bar K^*][0(1^+)] \to \bar K^* \bar K \pi$ decays. Here, $T_{ss}$ stands for the $I(J^P)=0(1^+)$ $\bar K \bar K^*$  and $I(J^P)=0(1^+)$ $\bar K^* \bar K^*$ molecular states in (a) and (b), respectively.}\label{fig:Schematicdecay1}
\end{figure}

The allowed strong decay channels of the $I(J^P)=0(1^+)$ $\bar K \bar K^*$ molecular state are $\bar{K}\bar{K}\pi$ and $\bar{K}\bar{K}\pi\pi$. Among them, the width of the four-body decay channel $\bar{K}\bar{K}\pi\pi$ is expected to be small because of its small phase space. In the following, we employ the effective Lagrangian approach to estimate the decay width of the $\bar K \bar K \pi$ channel for the $I(J^P)=0(1^+)$ $\bar K \bar K^*$ molecular state, which is similar to the calculation of the decay width of the $D D \pi$ channel for the $T_{cc}(3875)^+$ in the tree-level diagram in Refs. \cite{Meng:2021jnw,Ling:2021bir,Fleming:2021wmk,Feijoo:2021ppq}. The diagram for depicting the $[\bar K \bar K^*][0(1^+)] \to \bar K \bar K \pi$ decay is illustrated in Fig.~\ref{fig:Schematicdecay1}(a). The effective Lagrangian describing the interaction between the $I(J^P)=0(1^+)$ $\bar K \bar K^*$ molecular state and the constituent hadrons $\bar{K}$ and $\bar{K}^*$ can be constructed as follows:
\begin{eqnarray}
    \mathcal{L}_{T_{ss}\bar K \bar K^*} & = g_{T_{ss}\bar K \bar K^*}T_{ss}^\mu \widetilde K_\mu^* \widetilde K.
\end{eqnarray}
Here, $T_{ss}$ is the $I(J^P)=0(1^+)$ $\bar K \bar K^*$ molecular state, and the coupling constant $g_{T_{ss} \bar K \bar K^*}$ can be evaluated using the Weinberg's composite relation \cite{Weinberg:1962hj}, which allows us to obtain
\begin{eqnarray}
    g_{T_{ss}\bar K \bar K^*}^2 & = \dfrac{16\pi(m_{\bar K}+m_{\bar K^*})^{5/2}\sqrt{2|E_B|}}{(m_{\bar K} m_{\bar K^*})^{1/2}},
\end{eqnarray}
where $E_B$ is the binding energy of the $I(J^P)=0(1^+)$ $\bar K \bar K^*$ molecular state. Based on the aforementioned effective Lagrangians, the scattering amplitude of the $[\bar K \bar K^*][0(1^+)] \to \bar K \bar K \pi$ process is given by
\begin{eqnarray}
    i\mathcal{M}_{[\bar{K}\bar{K}^*][0(1^+)]\to\bar{K}\bar{K}\pi}&=& i g_{T_{ss}\bar K \bar K^*}\epsilon_{T_{ss}}^\lambda \frac{i(-g_{\lambda\rho}+q_\lambda q_\rho/m_{\bar K^*}^2)}{q^2-m_{\bar K^*}^2+im_{\bar K^*}\Gamma_{\bar K^*}}\nonumber\\
    &\times&\sqrt{m_{\bar K^*}m_{\bar K}}\left(\frac{-2g'}{f_\pi}\right)(ip_2^\rho).
\end{eqnarray}
The three-body partial decay width is defined by the following formula \cite{ParticleDataGroup:2022pth}
\begin{eqnarray}
    \Gamma &=& 2\frac{1}{\mathcal{S}}\frac{1}{2m_{T_{ss}}}\sum_{{\rm pol.}}\frac{1}{3}\int |\mathcal{M}|^2 d\Phi_3\nonumber\\
           &=& 2\frac{1}{\mathcal{S}}\frac{1}{(2\pi)^3} \frac{1}{32m_{T_{ss}}^3}\sum_{{\rm pol.}}\frac{1}{3} \int |\mathcal{M}|^2 dm_{12}^2dm_{23}^2,\label{threebodydecaywidth}
\end{eqnarray}
where $m_{12}^2=(p_1+p_2)^2$, $m_{23}^2=(p_2+p_3)^2$, and the factor 2 arises from the sum of two different decay final states $K^{-}K^-\pi^+$ and $\bar{K}^{0}\bar{K}^0\pi^-$. $\mathcal{S}$ is the statistical factor, we can obtain $\mathcal{S}=2$ when two identical $\bar K$ mesons appear in the final states, and $\mathcal{S}=1$ when no identical particle appears in the final states. Given that the $I(J^P)=0(1^+)$ $\bar K \bar K^*$ molecular state has not been observed, we take three representative binding energies of the $I(J^P)=0(1^+)$ $\bar K \bar K^*$ molecular state to estimate the decay width of the $[\bar K \bar K^*][0(1^+)] \to \bar K \bar K \pi$ process, and the corresponding numerical results are presented in the following:
\begin{eqnarray}
\renewcommand\tabcolsep{2.10cm}
\renewcommand{\arraystretch}{1.50}
\begin{array}{*{4}c}
\hline
E_B\,(\mathrm{MeV})     &~~~~~-4     &~~~~~-17    &~~~~~-30 \nonumber\\
\Gamma_{[\bar{K}\bar{K}^*][0(1^+)]\to\bar{K}\bar{K}\pi}\,(\mathrm{MeV}) &~~~~~37.0  &~~~~~43.2  &~~~~~33.8  \\
\hline
\end{array}.
\end{eqnarray}
Here, we need to mention that the decay width of the $[\bar K \bar K^*][0(1^+)] \to \bar K \bar K \pi$ process is approximately $16.7\,{\rm MeV}$ when the binding energy of the $I(J^P)=0(1^+)$ $\bar K \bar K^*$ molecular state is taken to be $-60\,{\rm MeV}$. Our obtained results is particularly close to the numerical result predicted by the quark-pair creation model in Ref. \cite{Ji:2024znj}, where the decay width of the $[KK^*][0(1^+)] \to KK \pi$ process is around $17\,{\rm MeV}$ when the binding energy of the $I(J^P)=0(1^+)$ $KK^*$ molecular state is taken to be $-60\,{\rm MeV}$. It can thus be concluded that the $\bar K \bar K \pi$ is a significant three-body strong decay channel for the $I(J^P)=0(1^+)$ $\bar K \bar K^*$ molecular state. In light of the observation of the double-charm tetraquark state $T_{cc}(3875)^+$ in the $D^0D^0\pi^+$ invariant mass spectrum \cite{LHCb:2021vvq}, it is recommended that the future experiments should prioritize the $\bar K \bar K \pi$ invariant mass spectrum when searching for the $I(J^P)=0(1^+)$ $\bar K \bar K^*$ molecular state, since the $I(J^P)=0(1^+)$ $\bar K \bar K^*$ state and the $T_{cc}(3875)^+$ state exhibit similar structures within the molecular framework.

\begin{figure}[htbp]
    \centering
    \includegraphics[width=8.0cm]{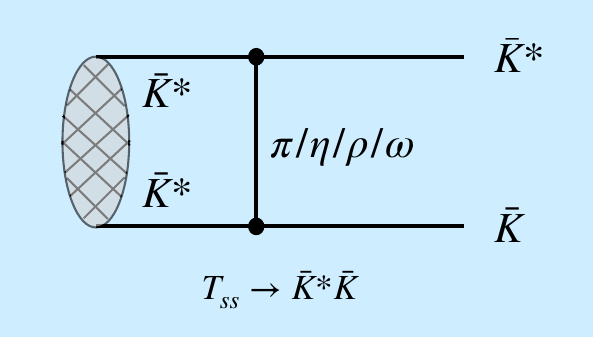}
    \caption{The diagram for depicting the $[\bar K^* \bar K^*][0(1^+)] \to \bar K^* \bar K$ decay. Here, $T_{ss}$ is the $I(J^P)=0(1^+)$ $\bar K^* \bar K^*$ molecular state.}\label{fig:Schematicdecay3}
\end{figure}

For the $I(J^P)=0(1^+)$ $\bar K^* \bar K^*$ molecular state, the allowed strong decay channels include $\bar{K}^*\bar{K}$, $\bar{K}^*\bar{K}\pi$, $\bar{K}\bar{K}\pi$, and $\bar{K}\bar{K}\pi\pi$. Among these decay channels, the four-body decay channel $\bar{K}\bar{K}\pi\pi$ should be small due to the presence of $P$-wave coupling and phase space suppression, while the three-body decay channel $\bar{K}\bar{K}\pi$ may have considerable contribution from the subsequent decay process $[\bar{K^*}\bar{K}^*][0(1^+)]\to\bar{K}^*\bar{K}\to\bar{K}\bar{K}\pi$. Thus, the $\bar{K}\bar{K}^*$ and $\bar{K}\bar{K}^*\pi$ channels should dominant the strong decays of the $I(J^P)=0(1^+)$ $\bar K^* \bar K^*$ molecular state, of which the $\bar K \bar K^*$ channel is the $S$-wave coupling and the $\bar K \bar K^* \pi$ channel is the $P$-wave coupling. First, we employ the effective Lagrangian approach to estimate the decay width of the $\bar K \bar K^*$ channel for the $I(J^P)=0(1^+)$ $\bar K^* \bar K^*$ molecular state in the tree-level diagram (see Fig.~\ref{fig:Schematicdecay3}), which has been widely applied to the study of two-body decays of hadronic molecules \cite{Zhang:2006ix,Chen:2017xat,Chen:2021tip,Luo:2023hnp,Chen:2024xlw,Zhang:2024usz}. For the $I(J^P)=0(1^+)$ $\bar K^* \bar K^*$ molecular state decaying into the $\bar{K}^*\bar{K}$ process, the decay amplitude ${\cal M}_{[AB]\to CD}$ is related to the scattering amplitude ${\cal M}_{AB\to CD}({\bf k},{\bf p})$ by the following relation \cite{Zhang:2006ix}
\begin{equation}
{\cal M}_{[AB]\to CD}=\sqrt{\frac{m_{[AB]}}{2m_Am_B}}\int{\rm d}^3{\bf k}\frac{{\rm d}^3 {\bf r}}{(2\pi)^3}\psi_{[AB]}({\bf r}){\cal M}_{AB\to CD}({\bf k},{\bf p}),
\end{equation}
where $\psi_{[AB]}({\bf r})$ is the spatial wave function of the hadronic molecular state $[AB]$, while ${\bf k}$ and ${\bf p}$ are the momenta of hadrons $A$ and $C$ in the center-of-mass frame, respectively. Then the two-body partial decay width can be calculated as follows \cite{ParticleDataGroup:2022pth}:
\begin{equation}
\Gamma_{[AB]\to CD}=\frac{p}{32\pi^2m_{[AB]}^2}\int{\rm d}\Omega|{\cal M}_{[AB]\to CD}|^2.
\end{equation}
In this work, both the spatial wave function $\psi_{[\bar{K}^*\bar{K}^*]}({\bf r})$ and the scattering amplitude ${\cal M}_{\bar{K}^*\bar{K}^* \to \bar{K}^*\bar{K}}$ have been obtained from the OBE model. In the realistic calculations, the decay width of the $[\bar K^* \bar K^*][0(1^+)] \to \bar K^* \bar K$ process is dependent on the cutoff $\Lambda$ in the form factor and the binding energy of the $I(J^P)=0(1^+)$ $\bar K^* \bar K^*$ molecular state. Nevertheless, there are currently no relevant experimental information that can be used to scale the cutoff value $\Lambda$ and the binding energy of the $I(J^P)=0(1^+)$ $\bar K^* \bar K^*$ molecular state exactly. In the numerical analysis, we take a typical value $\Lambda=1\,{\rm GeV}$ in the form factor \cite{Chen:2016qju} and three representative binding energies of the $I(J^P)=0(1^+)$ $\bar K^* \bar K^*$ molecular state to simply estimate the decay width of the $[\bar K^* \bar K^*][0(1^+)] \to \bar K^* \bar K$ process. The resulting numerical results are presented in the following:
\begin{eqnarray}
\renewcommand\tabcolsep{2.10cm}
\renewcommand{\arraystretch}{1.50}
\begin{array}{*{4}c}
\hline
E_B\,(\mathrm{MeV})     &~~~~~-4     &~~~~~-17    &~~~~~-30 \nonumber\\
\Gamma_{[\bar{K}^*\bar{K}^*][0(1^+)]\to\bar{K}^*\bar{K}}\,(\mathrm{MeV}) &~~~~~128.5  &~~~~~256.6  &~~~~~342.2  \\
\hline
\end{array}.
\end{eqnarray}
Here, we need to indicate that the decay width of the $[\bar K^* \bar K^*][0(1^+)] \to \bar K^* \bar K$ process depends on the input parameters. Nevertheless, the decay width of the $[\bar K^* \bar K^*][0(1^+)] \to \bar K^* \bar K$ process remains approximately within the same order of magnitude, at around hundreds of MeV, when the binding energy of the $I(J^P)=0(1^+)$ $\bar K^* \bar K^*$ molecular state varies between $-4\,\mathrm{MeV}$ and $-30\,\mathrm{MeV}$. Thus, the conclusion is that the $\bar K \bar K^*$ is the important two-body strong decay mode for the $I(J^P)=0(1^+)$ $\bar K^* \bar K^*$ molecular state. Furthermore, it is apparent that the $[\bar K^* \bar K^*][0(1^+)] \to \bar K^* \bar K$ process displays a relatively substantial decay width. However, if we refer the strong decay behaviors of the light-flavor hadrons as documented in the Particle Data Group \cite{ParticleDataGroup:2022pth}, we find that there are numerous light-flavor hadrons with decay widths of approximately hundreds of MeV. Therefore, the decay width of the $[\bar K^* \bar K^*][0(1^+)] \to \bar K^* \bar K$ process can be compared to that of other light-flavor hadrons \cite{ParticleDataGroup:2022pth}. In addition, the theoretical predictions of two-body strong decay widths for some light-flavor hadronic molecular states can also reach hundreds of MeV, such as the hidden-strange molecular pentaquarks $N(1875)/N(2270)$ \cite{Lin:2018kcc,Ben:2024qeg}, the hidden-strange molecular hexaquarks  $\Lambda \bar \Lambda$ \cite{Dong:2017rmg}, the dynamically generated states of two vector mesons $f_2(1270)/f_0(1500)$ \cite{Shen:2024jfr}, and so on.

Subsequently, we focus on the decay width of the $\bar K^* \bar K \pi$ channel for the $I(J^P)=0(1^+)$ $\bar K^* \bar K^*$ molecular state. Similar to the study of the decay width of the $[\bar K \bar K^*][0(1^+)] \to \bar K \bar K \pi$ process, we can estimate the decay width of the $[\bar K^* \bar K^*][0(1^+)] \to \bar K^* \bar K \pi$ process. The diagram for depicting the $[\bar K^* \bar K^*][0(1^+)] \to \bar K^* \bar K \pi$ decay is shown in Fig.~\ref{fig:Schematicdecay1}(b). The interaction between the $I(J^P)=0(1^+)$ $\bar K^* \bar K^*$ molecular state and the constituent hadrons $\bar{K}^*$ and $\bar{K}^*$ is described by the following effective Lagrangian
\begin{eqnarray}
    \mathcal{L}_{T_{ss}\bar K^*\bar K^*} = g_{T_{ss}\bar K^*\bar K^*} \epsilon_{\mu\nu\lambda\rho}\partial^\mu T_{ss}^{\nu} \widetilde K^{*\lambda} \widetilde K^{*\rho}.
\end{eqnarray}
Here, $T_{ss}$ is the $I(J^P)=0(1^+)$ $\bar K^* \bar K^*$ molecular state, we can evaluate the coupling constant $g_{T_{ss}\bar K^*\bar K^*}$ by means of the Weinberg's composite relation \cite{Weinberg:1962hj}, i.e.,
\begin{eqnarray}
    g_{T_{ss}\bar K^*\bar K^*}^2 = \frac{16\pi\, \sqrt{|E_B|}}{m_{\bar K^*}^{1/2}},
\end{eqnarray}
where $E_B$ is the binding energy of the $I(J^P)=0(1^+)$ $\bar K^* \bar K^*$ molecular state. The scattering amplitude of the $[\bar K^* \bar K^*][0(1^+)] \to \bar K^* \bar K \pi$ process is given by
\begin{eqnarray}
    i\mathcal{M}_{[\bar{K}^*\bar{K}^*][0(1^+)]\to\bar{K}^*\bar{K}\pi} &=& i g_{T_{ss}\bar K^*\bar K^*}\epsilon_{\alpha\beta\lambda\rho}(-ip^\alpha)\epsilon_{T_{ss}}^{\beta}\epsilon_{K^*}^{*\rho}g^{\lambda\theta}\nonumber\\
    &\times&\frac{i(-g_{\theta\tau}+q_\theta q_\tau/m_{\bar K^*})}{q^2-m_{\bar K^*}^2+im_{\bar K^*}\Gamma_{\bar K^*}}\sqrt{m_{\bar K} m_{\bar K^*}}\nonumber\\
    &\times&\left(\frac{-2g'}{f_\pi}\right)(ip_2^\tau).
\end{eqnarray}
By employing the formula presented in Eq. (\ref{threebodydecaywidth}), we can calculate the decay width of the $[\bar K^* \bar K^*][0(1^+)] \to \bar K^* \bar K \pi$ process. In the following, we list the obtained decay width of the $[\bar K^* \bar K^*][0(1^+)] \to \bar K^* \bar K \pi$ process when taking three representative binding energies of the $I(J^P)=0(1^+)$ $\bar K^* \bar K^*$ molecular state:
\begin{eqnarray}
\renewcommand\tabcolsep{2.10cm}
\renewcommand{\arraystretch}{1.50}
\begin{array}{*{4}c}
\hline
E_B\,(\mathrm{MeV})     &~~~~~-4     &~~~~~-17    &~~~~~-30 \nonumber\\
\Gamma_{[\bar{K}^*\bar{K}^*][0(1^+)]\to\bar{K}^*\bar{K}\pi}\,(\mathrm{MeV}) &~~~~~39.2  &~~~~~51.7  &~~~~~45.4  \\
\hline
\end{array}.
\end{eqnarray}
Obviously, the $\bar K^* \bar K \pi$ channel exist significant decay width for the $I(J^P)=0(1^+)$ $\bar K^* \bar K^*$ molecular state. However, the decay width for the $\bar K^* \bar K \pi$ channel is relatively small compared to the $\bar K^* \bar K$ channel, which is due to the $\bar K^* \bar K \pi$ channel is three-body and $P$-wave decay and the $\bar K^* \bar K$ channel is two-body and $S$-wave decay for the $I(J^P)=0(1^+)$ $\bar K^* \bar K^*$ molecular state.

By analyzing the strong decay behaviors of the $I(J^P)=0(1^+)$ $\bar K \bar K^*$ and $I(J^P)=0(1^+)$ $\bar K^* \bar K^*$ molecular states, we find that several channels exhibit significant decay widths, which have the potential to observe the $I(J^P)=0(1^+)$ $\bar K \bar K^*$ and $I(J^P)=0(1^+)$ $\bar K^* \bar K^*$ molecular states in future experiments. Before closing this subsection, we need to mention that the experimental data on the hadronic molecular state decays are not yet sufficient to be compared with theoretical results, due to the lack of the experimental results. Thus, it is challenging to make comprehensive and reliable predictions regarding the strong decay behaviors of the hadronic molecular states, which are sensitive to the input parameters, such as the binding energy, the cutoff parameter, and so on. The aim of the present discussion is to provide some significant strong decay channels for the $I(J^P)=0(1^+)$ $\bar K \bar K^*$ and $I(J^P)=0(1^+)$ $\bar K^* \bar K^*$ molecular candidates, which is essential for the experimental search for the double-strangeness molecular tetraquarks. It is recommended that the future theoretical studies of the strong decay behaviors of the $\bar K^{(*)} \bar K^{(*)}$ molecular tetraquark candidates can be conducted using others approaches and methods, with the objective of enhancing our understanding of these strong decay behaviors.

\subsection{Magnetic moments and M1 radiative decay widths of the $I(J^P)=0(1^+)$ $\bar K \bar K^*$ and $I(J^P)=0(1^+)$ $\bar K^* \bar K^*$ molecular candidates}

As a pivotal aspect of spectroscopic properties, another topic of the present work is to explore the magnetic moments and M1 radiative decay widths of the $\bar K^{(*)} \bar K^{(*)}$ molecular candidates, which are the important properties to reveal their inner structures. In practice, we employ the constituent quark model, which is a popular approach for quantifying the magnetic moments and M1 radiative decay widths of the hadronic states \cite{Liu:2003ab,Huang:2004tn,Zhu:2004xa,Haghpayma:2006hu,Wang:2016dzu,Deng:2021gnb,Gao:2021hmv,Zhou:2022gra,Wang:2022tib,Li:2021ryu,Schlumpf:1992vq,Schlumpf:1993rm,Cheng:1997kr,Ha:1998gf,Ramalho:2009gk,Girdhar:2015gsa,Menapara:2022ksj,Mutuk:2021epz,Menapara:2021vug,Menapara:2021dzi,Gandhi:2018lez,Dahiya:2018ahb,Kaur:2016kan,Thakkar:2016sog,Shah:2016vmd,Dhir:2013nka,Sharma:2012jqz,Majethiya:2011ry,Sharma:2010vv,Dhir:2009ax,Simonis:2018rld,Ghalenovi:2014swa,Kumar:2005ei,Rahmani:2020pol,Hazra:2021lpa,Gandhi:2019bju,Majethiya:2009vx,Shah:2016nxi,Shah:2018bnr,Ghalenovi:2018fxh,Wang:2022nqs,Mohan:2022sxm,An:2022qpt,Kakadiya:2022pin,Wu:2022gie,Wang:2023bek,Wang:2023aob,Wang:2023ael,Lai:2024jfe,Guo:2023fih,Li:2024wxr,Li:2024jlq,Sheng:2024hkf,Wang:2024sbw,Lei:2023ttd}.

Within the constituent quark model, the magnetic moments of the $\bar K^{(*)} \bar K^{(*)}$ molecular candidates, designated as $\mu_{T_{ss}}$, can be calculated using  the following equation \cite{Liu:2003ab,Huang:2004tn,Zhu:2004xa,Haghpayma:2006hu,Wang:2016dzu,Deng:2021gnb,Gao:2021hmv,Zhou:2022gra,Wang:2022tib,Li:2021ryu,Schlumpf:1992vq,Schlumpf:1993rm,Cheng:1997kr,Ha:1998gf,Ramalho:2009gk,Girdhar:2015gsa,Menapara:2022ksj,Mutuk:2021epz,Menapara:2021vug,Menapara:2021dzi,Gandhi:2018lez,Dahiya:2018ahb,Kaur:2016kan,Thakkar:2016sog,Shah:2016vmd,Dhir:2013nka,Sharma:2012jqz,Majethiya:2011ry,Sharma:2010vv,Dhir:2009ax,Simonis:2018rld,Ghalenovi:2014swa,Kumar:2005ei,Rahmani:2020pol,Hazra:2021lpa,Gandhi:2019bju,Majethiya:2009vx,Shah:2016nxi,Shah:2018bnr,Ghalenovi:2018fxh,Wang:2022nqs,Mohan:2022sxm,An:2022qpt,Kakadiya:2022pin,Wu:2022gie,Wang:2023bek,Wang:2023aob,Wang:2023ael,Lai:2024jfe,Guo:2023fih,Li:2024wxr,Li:2024jlq,Sheng:2024hkf,Wang:2024sbw,Lei:2023ttd}
\begin{eqnarray}\label{magneticmoment}
\mu_{T_{ss}}=\left\langle{J_{T_{ss}},J_{T_{ss}}\left|\sum_{j}\hat{\mu}_{zj}^{\rm S}+\hat{\mu}_z^{\rm L}\right|J_{T_{ss}},J_{T_{ss}}}\right\rangle.
\end{eqnarray}
Here, the $z$-component of the spin magnetic moment operator of the $j$-th constituent of the hadron, represented by $\hat{\mu}_{zj}^{\rm S}$, can be expressed as follows \cite{Liu:2003ab,Huang:2004tn,Zhu:2004xa,Haghpayma:2006hu,Wang:2016dzu,Deng:2021gnb,Gao:2021hmv,Zhou:2022gra,Wang:2022tib,Li:2021ryu,Schlumpf:1992vq,Schlumpf:1993rm,Cheng:1997kr,Ha:1998gf,Ramalho:2009gk,Girdhar:2015gsa,Menapara:2022ksj,Mutuk:2021epz,Menapara:2021vug,Menapara:2021dzi,Gandhi:2018lez,Dahiya:2018ahb,Kaur:2016kan,Thakkar:2016sog,Shah:2016vmd,Dhir:2013nka,Sharma:2012jqz,Majethiya:2011ry,Sharma:2010vv,Dhir:2009ax,Simonis:2018rld,Ghalenovi:2014swa,Kumar:2005ei,Rahmani:2020pol,Hazra:2021lpa,Gandhi:2019bju,Majethiya:2009vx,Shah:2016nxi,Shah:2018bnr,Ghalenovi:2018fxh,Wang:2022nqs,Mohan:2022sxm,An:2022qpt,Kakadiya:2022pin,Wu:2022gie,Wang:2023bek,Wang:2023aob,Wang:2023ael,Lai:2024jfe,Guo:2023fih,Li:2024wxr,Li:2024jlq,Sheng:2024hkf,Wang:2024sbw,Lei:2023ttd}:
\begin{eqnarray}
\hat{\mu}_{zj}^{\rm S}=\mu_j\hat{\sigma}_{zj}~~~~~{\rm with}~~~~~\mu_j=\frac{e_j}{2m_j},
\end{eqnarray}
where the $z$-component of the Pauli spin vector operator of the $j$-th constituent of the hadron is represented by $\hat{\sigma}_{zj}$. The symbols $e_j$ and $m_j$ are the charge and mass of the $j$-th constituent of the hadron, respectively. Furthermore, the $z$-component of the orbital magnetic moment operator, represented by $\hat{\mu}_z^{\rm L}$, is given by the following equation \cite{Wang:2022nqs,Wang:2023aob,Wang:2023ael,Wang:2023bek,Lai:2024jfe,Sheng:2024hkf,Wang:2024sbw}:
\begin{eqnarray}
\hat{\mu}_z^{\rm L}=\mu_{\alpha\beta}^{\rm L}\hat{L}_z~~~~{\rm with}~~~~\mu_{\alpha\beta}^{\rm L}=\frac{m_{\alpha}}{m_{\alpha}+m_{\beta}}\mu_{\beta}+\frac{m_{\beta}}{m_{\alpha}+m_{\beta}}\mu_{\alpha},
\end{eqnarray}
where $\alpha$ and $\beta$ are the constituent hadrons of the $\bar K^{(*)} \bar K^{(*)}$ molecular candidates, and the $z$-component of the orbital angular momentum vector operator between the constituent hadrons is $\hat{L}_z$.

Now, let us turn to the  M1 radiative decay widths of the $\bar K^{(*)} \bar K^{(*)}$ molecular candidates. The constituent quark model enables us to calculate the transition magnetic moments between the $\bar K^{(*)} \bar K^{(*)}$ molecular candidates, denoted as $\mu_{{T_{ss}} \to {T_{ss}^{\prime}}}$, which is achieved by the relation \cite{Wang:2022nqs,Wang:2023aob,Wang:2023ael,Wang:2023bek,Lai:2024jfe,Sheng:2024hkf,Wang:2024sbw}:
\begin{eqnarray}\label{transitionmagneticmoment}
\mu_{{T_{ss}} \to {T_{ss}^{\prime}}}=\left\langle{J_{{T_{ss}^{\prime}}},J_{z}\left|\sum_{j}\hat{\mu}_{zj}^{\rm S}e^{-i {\bf q}\cdot{\bf r}_j}\right|J_{{T_{ss}}},J_{z}}\right\rangle.
\end{eqnarray}
Here, we take the condition $J_z={\rm Min}\{J_{T_{ss}},\,J_{{T_{ss}^{\prime}}}\}$. The spatial wave function of the emitted photon, represented by the expression $e^{-i {\bf q}\cdot{\bf r}_j}$, is a function of the momentum of the emitted photon ${\bf q}$ and the coordinate of the $j$-th quark ${\bf r}_j$ with
\begin{eqnarray}
q=\frac{m_{{T_{ss}}}^2-m_{{T_{ss}^{\prime}}}^2}{2m_{{T_{ss}}}}.
\end{eqnarray}
In the realistic calculation, the spherical Bessel function, $j_l(x)$, and the spherical harmonic function, $Y_{l m}(\Omega_{\bf x})$, are used to expand the expression $e^{-i{\bf q}\cdot{\bf r}_j}$ by the following relation \cite{Khersonskii:1988krb}:
\begin{eqnarray}
e^{-i{\bf q}\cdot{\bf r}_j}&=&4\pi\sum\limits_{l=0}^\infty\sum\limits_{m=-l}^l(-i)^lj_l(qr_j)Y_{lm}^*(\Omega_{\bf q})Y_{lm}(\Omega_{{\bf r}_j}).
\end{eqnarray}
With the above preparation, the M1 radiative decay width between the $\bar K^{(*)} \bar K^{(*)}$ molecular candidates with $J=1$, designated as $\Gamma_{{T_{ss}} \to {T_{ss}^{\prime}}\gamma}$, is given by the following relation \cite{Wang:2022nqs,Wang:2023aob,Wang:2023ael,Wang:2023bek,Lai:2024jfe,Sheng:2024hkf,Wang:2024sbw}:
\begin{eqnarray}
 \Gamma_{{T_{ss}} \to {T_{ss}^{\prime}}\gamma}=\alpha_{\rm {EM}}\frac{2}{3}\frac{q^{3}}{m_{p}^{2}} \frac{\left|\mu_{{T_{ss}} \to {T_{ss}^{\prime}}}\right|^2}{\mu_N^2},
\end{eqnarray}
where $\alpha_{\rm {EM}} \approx {1}/{137}$, $m_p$, and $\mu_N=e/2m_p$ are the electromagnetic fine structure constant, the mass of the proton, and the nuclear magneton, respectively.

To calculate the overlap of the spatial wave functions of the initial and final states in Eqs.~(\ref{magneticmoment}) and (\ref{transitionmagneticmoment}), we utilise the  numerical spatial wave functions of the $\bar K^{(*)} \bar K^{(*)}$ molecular candidates, which are derived by studying their mass spectra. Furthermore, the simple harmonic oscillator wave function is employed to delineate the spatial wave functions of the strange mesons $\bar K$ and $\bar K^*$ \cite{Wang:2022tib,Zhou:2022gra,Wang:2022nqs,Wang:2023aob,Wang:2023ael,Wang:2023bek,Lai:2024jfe,Sheng:2024hkf,Wang:2024sbw}. For the $\bar K$ and $\bar K^*$, the radial quantum number $n$, the orbital quantum number $l$, and the magnetic quantum number $m$ are $0$.  In this case, the simple harmonic oscillator wave function $\phi(\beta,{\bf r})$ can be explicitly expressed as
\begin{eqnarray}
\phi(\beta,{\bf r})=\frac{\beta^{{3}/{2}}{\mathrm e}^{-{\beta^2r^2}/{2}}}{\pi^{{3}/{4}}}.
\end{eqnarray}
Here, $\beta$ represents a phenomenological parameter with the values of $\beta_{K}=0.46~{\rm GeV}$ and $\beta_{K^*}=0.32~{\rm GeV}$ \cite{Close:2005se}.
In the present work, both the form factor and the spatial wave function of the meson can reflect that the mesons are not point-like particles. When deriving the OBE effective interactions between hadrons, the form factor depicts the structure of a virtual particle in the off-shell, which is a straightforward extension of the traditional meson exchange model involved in the nuclear force \cite{Yukawa:1935xg}. This effective approach has been successfully employed to predict a series of hadronic molecular states, and reproduce the bound-state properties of the observed $P_{c}(4312)$, $P_{c}(4440)$, $P_{c}(4457)$ \cite{Aaij:2019vzc}, and $T_{cc}(3875)^+$ \cite{LHCb:2021vvq} under the hadronic molecular picture. When calculating the magnetic moments and M1 radiative decay widths of hadrons, the spatial wave function of the meson represents an effective degree of freedom, which describes the distribution of quarks within the meson. This formalism has been extensively used to study the radiative decay widths and magnetic moments of hadronic states \cite{Liu:2003ab,Huang:2004tn,Zhu:2004xa,Haghpayma:2006hu,Wang:2016dzu,Deng:2021gnb,Gao:2021hmv,Zhou:2022gra,Wang:2022tib,Li:2021ryu,Schlumpf:1992vq,Schlumpf:1993rm,Cheng:1997kr,Ha:1998gf,Ramalho:2009gk,Girdhar:2015gsa,Menapara:2022ksj,Mutuk:2021epz,Menapara:2021vug,Menapara:2021dzi,Gandhi:2018lez,Dahiya:2018ahb,Kaur:2016kan,Thakkar:2016sog,Shah:2016vmd,Dhir:2013nka,Sharma:2012jqz,Majethiya:2011ry,Sharma:2010vv,Dhir:2009ax,Simonis:2018rld,Ghalenovi:2014swa,Kumar:2005ei,Rahmani:2020pol,Hazra:2021lpa,Gandhi:2019bju,Majethiya:2009vx,Shah:2016nxi,Shah:2018bnr,Ghalenovi:2018fxh,Wang:2022nqs,Mohan:2022sxm,An:2022qpt,Kakadiya:2022pin,Wu:2022gie,Wang:2023bek,Wang:2023aob,Wang:2023ael,Lai:2024jfe,Guo:2023fih,Li:2024wxr,Li:2024jlq,Sheng:2024hkf,Wang:2024sbw,Lei:2023ttd}, and reproduce experimental data on the magnetic moments of the decuplet and octet baryons quantitatively \cite{Schlumpf:1993rm, Kumar:2005ei, Ramalho:2009gk}.

In the present work, the constituent quark masses are taken to be
\begin{eqnarray*}
m_u=336\,{\rm MeV},~~~m_d=336\,{\rm MeV},~~~{\rm and}~~~m_s=540\,{\rm MeV}
\end{eqnarray*}
from Ref. \cite{Kumar:2005ei} for quantifying the magnetic moments and M1 radiative decay widths of the $I(J^P)=0(1^+)$ $\bar K \bar K^*$ and $I(J^P)=0(1^+)$ $\bar K^* \bar K^*$ molecular states, which has been frequently employed in the discussion of the hadronic electromagnetic properties over the past few decades \cite{Li:2021ryu,Zhou:2022gra,Wang:2022tib,Wang:2022nqs,Wang:2023aob,Wang:2023ael,Lai:2024jfe,Sheng:2024hkf,Wang:2024sbw}.

\renewcommand\tabcolsep{0.27cm}
\renewcommand{\arraystretch}{1.50}
\begin{table}[!htbp]
  \caption{Magnetic moments of the $I(J^P)=0(1^+)$ $\bar K \bar K^*$ and $I(J^P)=0(1^+)$ $\bar K^* \bar K^*$ molecular states by considering three analytical scenarios: (I) the single-channel analysis, (II) the $S$-$D$ wave mixing analysis, and (III) the coupled-channel analysis. Here, we present the numerical results after including $S$-$D$ wave mixing and coupled-channel effects for three binding energies of $-4$, $-17$, and $-30~{\rm MeV}$.}
  \label{Magneticmoments}
\centering
\begin{tabular}{c|c|c}
\toprule[1.0pt]
\toprule[1.0pt]
$T_{ss}$ molecules&Scenarios&Magnetic moments \\\midrule[1.0pt]
\multirow{3}{*}{$[\bar K \bar K^*][0(1^+)]$}&I&$-1.16~\mu_N$\\
                                  &II&$-1.14~\mu_N$\,, $-1.14~\mu_N$\,, $-1.14~\mu_N$\\
                                 &III&$-1.16~\mu_N$\,, $-1.21~\mu_N$\,, $-1.23~\mu_N$\\\hline
\multirow{2}{*}{$[\bar K^* \bar K^*][0(1^+)]$}&I&$-1.16~\mu_N$\\
                                  &II&$-1.14~\mu_N$\,, $-1.14~\mu_N$\,, $-1.13~\mu_N$\\
\bottomrule[1.0pt]
\bottomrule[1.0pt]
\end{tabular}
\end{table}

In Table \ref{Magneticmoments}, we present the magnetic moments of the $I(J^P)=0(1^+)$ $\bar K \bar K^*$ and $I(J^P)=0(1^+)$ $\bar K^* \bar K^*$ molecular states by considering the single-channel, $S$-$D$ wave mixing, and coupled-channel analysis. Focusing on the $S$-wave components, the magnetic moments of the $I(J^P)=0(1^+)$ $\bar K \bar K^*$ and $I(J^P)=0(1^+)$ $\bar K^* \bar K^*$ molecular states are
\begin{eqnarray}
\mu_{[\bar K \bar K^*][0(1^+)]}&=&\frac{1}{2}\mu_{\bar K^{*0}}+\frac{1}{2}\mu_{ K^{*-}},\\
\mu_{[\bar K^* \bar K^*][0(1^+)]}&=&\frac{1}{2}\mu_{\bar K^{*0}}+\frac{1}{2}\mu_{ K^{*-}},
\end{eqnarray}
respectively. It can be seen that the magnetic moments of the $I(J^P)=0(1^+)$ $\bar K \bar K^*$ and $I(J^P)=0(1^+)$ $\bar K^* \bar K^*$ molecular states are identical, with a value of $-1.16~\mu_N$, which is consistent with the magnetic moment behavior for the $I(J^P)=0(1^+)$ $D D^*$ and $I(J^P)=0(1^+)$ $D^* D^*$ molecular states \cite{Lei:2023ttd}. And then, the impact of the $D$-wave channels for their magnetic moments is considered. In this case, the magnetic moments of the hadronic molecular states are contingent upon their numerical spatial wave functions \cite{Wang:2022nqs,Wang:2023aob,Wang:2023ael,Wang:2023bek,Lai:2024jfe,Sheng:2024hkf,Wang:2024sbw}, which in turn are contingent upon the corresponding bound energies. Due to the $I(J^P)=0(1^+)$ $\bar K \bar K^*$ and $I(J^P)=0(1^+)$ $\bar K^* \bar K^*$ molecular states have not been observed, we present the numerical results for three binding energies of $-4$, $-17$, and $-30~{\rm MeV}$ of the discussed molecular states. Nevertheless, $S$-$D$ wave mixing effect has a negligible influence for the magnetic moments of the $I(J^P)=0(1^+)$ $\bar K \bar K^*$  and $I(J^P)=0(1^+)$ $\bar K^* \bar K^*$ molecular states, with a change in the magnetic moments of less than $0.03~\mu_N$ upon the addition of the contribution of the $D$-wave channels. This phenomenon can be attributed to the fact that the $S$-wave channels contribute predominantly, with probabilities over 96\%, to the formation of the $I(J^P)=0(1^+)$ $\bar K \bar K^*$ and $I(J^P)=0(1^+)$ $\bar K^* \bar K^*$ bound states. For the $I(J^P)=0(1^+)$ $\bar K \bar K^*$ molecular state, we can further consider coupled-channel effect including the $[\bar K \bar K^*]|{}^3\mathbb{S}_{1}\rangle$, $[\bar K \bar K^*]|{}^3\mathbb{D}_{1}\rangle$, $[\bar K^* \bar K^*]|{}^3\mathbb{S}_{1}\rangle$, and $[\bar K^* \bar K^*]|{}^3\mathbb{D}_{1}\rangle$ channels. The incorporation of coupled-channel effect gives rise to modifications for the magnetic moment of the $I(J^P)=0(1^+)$ $\bar K \bar K^*$ molecular state, with a change of up to $0.07~\mu_N$.

\renewcommand\tabcolsep{0.03cm}
\renewcommand{\arraystretch}{1.50}
\begin{table}[!htbp]
  \caption{Transition magnetic moments and M1 radiative decay widths of the $[\bar K^* \bar K^*][0(1^+)] \to [\bar K \bar K^*][0(1^+)] \gamma$ process by considering three analytical scenarios: (I) the single-channel analysis, (II) the $S$-$D$ wave mixing analysis, and (III) the coupled-channel analysis. Here, we take three binding energies of $-4$, $-17$, and $-30~{\rm MeV}$ to present these numerical results.}
  \label{Radiativedecays}
\centering
\begin{tabular}{c|c|c}
\toprule[1.0pt]
\toprule[1.0pt]
Physical quantities&Scenarios&Values \\\midrule[1.0pt]
\multirow{3}{*}{$\mu_{[\bar K^* \bar K^*][0(1^+)] \to [\bar K \bar K^*][0(1^+)]}$}&I&$-0.09~\mu_N$\,, $-0.13~\mu_N$\,, $-0.14~\mu_N$\\
                                  &II&$-0.07~\mu_N$\,, $-0.12~\mu_N$\,, $-0.13~\mu_N$\\
                                 &III&$-0.17~\mu_N$\,, $-0.36~\mu_N$\,, $-0.43~\mu_N$\\\hline
\multirow{3}{*}{$\Gamma_{[\bar K^* \bar K^*][0(1^+)] \to [\bar K \bar K^*][0(1^+)]\gamma}$}&I&$2.11~{\rm keV}$\,, $4.06~{\rm keV}$\,, $4.62~{\rm keV}$\\
                                  &II&$1.20~{\rm keV}$\,, $3.36~{\rm keV}$\,, $3.98~{\rm keV}$\\
                                 &III&$7.04~{\rm keV}$\,, $30.91~{\rm keV}$\,, $44.54~{\rm keV}$\\
\bottomrule[1.0pt]
\bottomrule[1.0pt]
\end{tabular}
\end{table}

Table \ref{Radiativedecays} presents the transition magnetic moments and M1 radiative decay widths of the $[\bar K^* \bar K^*][0(1^+)] \to [\bar K \bar K^*][0(1^+)] \gamma$ process, where we take three binding energies of $-4$, $-17$, and $-30~{\rm MeV}$ to present these numerical results.  In the case of the single-channel analysis, the transition magnetic moment of the $[\bar K^* \bar K^*][0(1^+)] \to [\bar K \bar K^*][0(1^+)] \gamma$ process is in the range from $-0.09$  to $-0.14~\mu_N$, and the corresponding M1 radiative decay width is in the range from $2.11$ to $4.62~{\rm keV}$.
In the case of the $S$-$D$ wave mixing analysis, the transition magnetic moment of the $[\bar K^* \bar K^*][0(1^+)] \to [\bar K \bar K^*][0(1^+)] \gamma$ process is estimated to be $-0.07 \sim -0.13~\mu_N$, with the corresponding M1 radiative decay width calculated to be approximately $1.20 \sim 3.98~{\rm keV}$. A comparison of the numerical results for the single-channel analysis demonstrates that $S$-$D$ wave mixing effect plays a relatively minor role in determining  the transition magnetic moment and M1 radiative decay width of the $[\bar K^* \bar K^*][0(1^+)] \to [\bar K \bar K^*][0(1^+)] \gamma$ process, while the changes of the transition magnetic moment and M1 radiative decay width are less than $0.02~\mu_N$ and $1.00~{\rm keV}$, respectively. This is due to the fact that the components of the $D$-wave channels, with probabilities less than 4\%, have a small influence on the formation of the $I(J^P)=0(1^+)$ $\bar K \bar K^*$ and $I(J^P)=0(1^+)$ $\bar K^* \bar K^*$ bound states.
In the case of the coupled-channel analysis, the transition magnetic moment of the $[\bar K^* \bar K^*][0(1^+)] \to [\bar K \bar K^*][0(1^+)] \gamma$ process is between $-0.17$ and $-0.43~\mu_N$, with the corresponding M1 radiative decay width lying between $7.04$ and $44.54~{\rm keV}$. A comparison of the results obtained from the single-channel and coupled-channel analysis reveals that the transition magnetic moment and M1 radiative decay width can change by up to $0.30~\mu_N$ and $40.00~{\rm keV}$, respectively. It can thus be concluded that coupled-channel effect plays a significant role in mediating the transition magnetic moment and M1 radiative decay width of the $[\bar K^* \bar K^*][0(1^+)] \to [\bar K \bar K^*][0(1^+)] \gamma$ process. After considering coupled-channel effect, the transition magnetic moment of the $[\bar K^* \bar K^*][0(1^+)] \to [\bar K \bar K^*][0(1^+)] \gamma$ process is influenced by an obvious contribution arising from $\mu_{[\bar K^* \bar K^*]\left|{}^3\mathbb{S}_{1}\right\rangle \to [\bar K^* \bar K^*]\left|{}^3\mathbb{S}_{1}\right\rangle}$, except for $\mu_{[\bar K^* \bar K^*]\left|{}^3\mathbb{S}_{1}\right\rangle \to [\bar K \bar K^*]\left|{}^3\mathbb{S}_{1}\right\rangle}$. As the binding energy of the $I(J^P)=0(1^+)$ $\bar K \bar K^*$ bound state increase, the contribution of the $[\bar K^* \bar K^*]|{}^3\mathbb{S}_{1}\rangle$ channel becomes evident listed in Table \ref{KKstarboundstate}, this will result in an increase for the contribution of $\mu_{[\bar K^* \bar K^*]\left|{}^3\mathbb{S}_{1}\right\rangle \to [\bar K^* \bar K^*]\left|{}^3\mathbb{S}_{1}\right\rangle}$, which will obviously affect the transition magnetic moment of the $[\bar K^* \bar K^*][0(1^+)] \to [\bar K \bar K^*][0(1^+)] \gamma$ process. To illustrate, the following relations can be obtained:
 \begin{eqnarray}
 \mu_{[\bar K^* \bar K^*]\left|{}^3\mathbb{S}_{1}\right\rangle \to [\bar K \bar K^*]\left|{}^3\mathbb{S}_{1}\right\rangle}&=&-0.13~\mu_N,\\
 \mu_{[\bar K^* \bar K^*]\left|{}^3\mathbb{S}_{1}\right\rangle \to [\bar K^* \bar K^*]\left|{}^3\mathbb{S}_{1}\right\rangle}&=&-0.31~\mu_N,
 \end{eqnarray}
 when the binding energies of the $I(J^P)=0(1^+)$ $\bar K \bar K^*$ and $I(J^P)=0(1^+)$ $\bar K^* \bar K^*$ molecular states are taken to be $-30\,{\rm MeV}$ by incorporating coupled-channel effect.

By studying the strong and radiative decay behaviors of the $\bar K^{(*)} \bar K^{(*)}$ molecular tetraquark candidates, we see that the radiative decay widths of the $\bar K^{(*)} \bar K^{(*)}$ molecular tetraquark candidates are rather smaller compared to their strong decay widths, which is analogous to the situation of the decay behaviors between the processes of $K^* \to  K \gamma$ and $K^* \to  K \pi$ \cite{ParticleDataGroup:2022pth}.

\section{Summary and outlook}\label{sec4}

The study of hadronic molecular states in the light-quark sector has become a significant and intriguing topic in hadron spectroscopy over the past few decades. This research provides valuable insights into non-perturbative behavior of strong interaction. Given the current research on the $K^{(*)} \bar K^{(*)}$ molecular tetraquarks and the $H$-dibaryon, we propose that both experimental and theoretical efforts should focus on double-strangeness molecular tetraquark candidates. These candidates exhibit exotic flavor quantum number, specifically $ss\bar q \bar q$. In this study, we explore the spectroscopic properties of the $\bar K^{(*)} \bar K^{(*)}$ systems.

We carry out a detailed dynamical calculation of the effective interactions for the $\bar K^{(*)} \bar K^{(*)}$ systems, using the OBE model. Our numerical results show that the OBE effective interaction for the $I(J^P)=1(0^+)$ $\bar K \bar K$ state agrees with the lattice QCD calculation, and $\pi$ exchange plays a key role in the interactions of the $I(J^P)=0(1^+)$ $\bar K \bar K^*$ and $I(J^P)=0(1^+)$ $\bar K^* \bar K^*$ states. By applying the OBE model, we can search for loosely bound states in these systems and predict the mass spectra of double-strangeness molecular tetraquarks. Our analysis also includes both $S$-$D$ wave mixing and coupled-channel effects. According to our results, the most likely candidates for double-strangeness molecular tetraquarks are the $I(J^P)=0(1^+)$ $\bar K \bar K^*$ and $I(J^P)=0(1^+)$ $\bar K^* \bar K^*$ states.

In addition to the predictions of mass spectra, we explore the strong decay behaviors, magnetic moments, M1 radiative decay widths of the $I(J^P)=0(1^+)$ $\bar K \bar K^*$ and $I(J^P)=0(1^+)$ $\bar K^* \bar K^*$ molecular tetraquark candidates. Using the effective Lagrangian approach, we estimate the strong decay behaviors of the $I(J^P)=0(1^+)$ $\bar K \bar K^*$ and $I(J^P)=0(1^+)$ $\bar K^* \bar K^*$ molecular states. Our numerical results indicate that several channels exhibit significant decay widths, which have the potential to observe the $I(J^P)=0(1^+)$ $\bar K \bar K^*$ and $I(J^P)=0(1^+)$ $\bar K^* \bar K^*$ molecular states in future experiments. Meanwhile, we analyze their magnetic moments and M1 radiative decay widths based on the constituent quark model. Our findings suggest that the magnetic moments both molecular tetraquarks are approximately $-1.16~\mu_N$, and the M1 radiative decay width for the process $[\bar K^* \bar K^*][0(1^+)] \to [\bar K \bar K^*][0(1^+)] \gamma$ ranges from several keV to several tens of keV, depending on their binding energies.

We hope our experimental colleagues will focus on searching for the $I(J^P)=0(1^+)$ $\bar K \bar K^*$ and $I(J^P)=0(1^+)$ $\bar K^* \bar K^*$ molecular tetraquarks in the near future. These candidates could provide crucial insights for establishing the molecular tetraquark states in the light-quark sector. Experimental detection of these states is promising, especially through weak decays of $B$ meson in facilities like LHCb and Belle II.

Before concluding this work, we should engage in an extended discussion. It is interesting to highlight the attractive character of the $\bar K\bar K^*$ and $\bar K^* \bar K^*$ interactions, as they can yield intriguing results across various scenarios. For example, these interactions may facilitate the production of the multi-strange particles in the high-energy collisions of protons or ions \cite{Blume:2011sb,Shi:2016elm}. Additionally, the strong binding within the $\bar K\bar K^*$ or $\bar K^* \bar K^*$ systems can lead to the formation of multiple bound states, such as the $\bar K\bar K^* \bar K^*$, $\bar K^*\bar K^*\bar K^*$, and others. $\bar K\bar K^* \bar K^*$ and $\bar K^*\bar K^*\bar K^*$ are typical few-body systems that deserve further exploration. It is similar to the theoretical studies of the $DDD^*$ \cite{Wu:2021kbu}, $D D^*D^*$ and $D^*D^*D^*$ \cite{Luo:2021ggs} molecular systems when reporting the observation of the $T_{cc}(3875)^+$ by LHCb \cite{LHCb:2021vvq}.

\section*{ACKNOWLEDGMENTS}

F.L.W. would like to thank J. Z. Wang for very helpful discussions. This work is supported by the National Natural Science Foundation of China under Grant Nos. 12405097, 12335001, 12247155, and 12247101, National Key Research and Development Program of China under Contract No. 2020YFA0406400, the 111 Project under Grant No. B20063, the fundamental Research Funds for the Central Universities, and the project for top-notch innovative talents of Gansu province. F.L.W. is also supported by the China Postdoctoral Science Foundation under Grant No. 2022M721440.

\end{document}